\documentclass[11pt,a4paper]{article}
\usepackage{jcappub}
\usepackage{amsxtra}
\usepackage[english]{babel}
\usepackage{amsfonts}
\allowdisplaybreaks[4]

\begin{document}

\title{Oscillating solutions of the matter density contrast in Horndeski's theory}

\author{Jiro Matsumoto}
\affiliation{LeCosPA, National Taiwan University, Taipei, Taiwan 10617, \\
Institute of Physics, Kazan Federal University, Kremlevskaya Street 16a,
Kazan 420008, Russia}

\emailAdd{jmatsumoto@ntu.edu.tw}
\abstract{
Evolution of the large scale structure of the Universe is described by galaxy bias and the growth rate function of the matter density contrast. 
The behavior of the growth rate function, in general, depends on the types of dark energy or modified gravity theories, 
because they modify Poisson's equation for gravity by changing the Newtonian constant in the quasi-static solutions. 
Whereas, another mode of the matter density contrast ``oscillating mode" has not been studied well, although 
there is a possibility that it drastically changes the gravitational law. In this paper, we will 
construct the way to investigate oscillating solutions and show the concrete forms of the solutions by using the WKB approximation. 
}
	
\arxivnumber{1806.10454}

\maketitle
\section{Introduction \label{sec1}}
The accelerated expansion of the Universe was discovered by the observations of type Ia supernovae in 
late 1990s \cite{Riess:1998cb,Perlmutter:1998np}, and it is now also supported by the other observations:
cosmic microwave background (CMB) radiation \cite{Komatsu:2010fb,Ade:2013zuv,Ade:2015xua},
baryon acoustic oscillations (BAO)
\cite{Percival:2009xn,Blake:2011en,Beutler:2011hx,Cuesta:2015mqa,Delubac:2014aqe},
and so on. 
To realize the accelerated expansion in the homogeneous and isotropic Universe, it is necessary to introduce some energy components with the equation of 
state parameter $w=p / \rho$ which is less than $-1/3$ in the Einstein equations. 
Dark energy is a hypothetical energy which has such a property. The most famous 
model of dark energy is the cosmological constant $\Lambda$, and the $\Lambda$ cold dark matter ($\Lambda$CDM) model is known as 
the standard model in cosmology. The $\Lambda$CDM model is a simple model but it is consistent with almost
all of the astronomical observations. 
While, there are also the other candidates of dark energy; 
the quintessence model \cite{Peebles:1987ek,Ratra:1987rm,Chiba:1997ej,Zlatev:1998tr}, scalar-tensor theory \cite{Jordan:1955,Brans:1961sx,Fujii:2003pa,Maeda:1988ab}, 
$F(R)$ modified gravity \cite{Buchdahl:1983zz,Nojiri:2006ri,Sotiriou:2008rp,DeFelice:2010aj,Nojiri:2010wj}, 
and so on. 
Horndeski's theory \cite{Horndeski} is known as a general theory of scalar-tensor theory including quintessence model as a special case. 

At the background level, the quintessence model, scalar-tensor theory, and $F(R)$ modified gravity can describe same expansion history of the Universe 
as in the $\Lambda$CDM model (e.g. the way of reconstructing quintessence potential from the expanding rate can be seen in Refs.~\cite{Huterer:1998qv,Starobinsky:1998fr}). 
Therefore, it is necessary to consider perturbative behavior of the models in order to distinguish the models of dark energy. 
The matter density contrast defined by $\delta \equiv \delta \rho / \rho$ describes the perturbative behavior of the matter distribution, and 
the growth rate function $f=d \ln \delta /d \ln a$ is usually used as the indicator of the matter density contrast. 
The evolution of the growth rate function, in general, depends on the property of the dark energy. 
For example, it is known that there is a scale dependence in the growth rate function in $F(R)$ modified gravity \cite{Tsujikawa:2007gd,delaCruzDombriz:2008cp} 
and the Newtonian constant in Poisson's equation is modified in Horndeski's theory \cite{DeFelice:2011hq}. 
These results are obtained by assuming quasi-static approximation, which treats
time derivatives of the perturbation quantities as the same order as the expanding speed of the Universe. 
While, there is another mode of the matter density perturbation ``oscillating mode" \cite{Bamba:2011ih,Matsumoto:2013sba}, 
which is caused by extra degrees of freedom, e.g. perturbation of the scalar field $\phi$. 
Oscillating behavior of the perturbation quantity in Horndeski's theory is, for example, seen in Fig.~8 in Ref.~\cite{Kobayashi:2009wr}. 
Although it is natural to consider oscillating mode because scalar field perturbation usually has non-zero sound speed, it has not been enough studied. 
There are two reasons why it has not been enough studied; mixing between perturbation quantities has not been carefully considered, and 
second leading order terms are necessary to be taken into account for the oscillating mode. In this paper, 
we will construct the method to obtain oscillating solutions in Horndeski's theory. 
The contents of the paper are as follows:
general background equations and perturbation equations in Horndeski's theory are given in Sec.~\ref{sec2}, 
oscillating solutions of the matter density contrast are derived in Sec.~\ref{sec3}, 
some examples of quasi-static solutions and oscillating solutions are expressed in Sec.~\ref{sec4},
and concluding remarks are in Sec.~\ref{sec5}.
We use natural units, $\hbar =c = k_\mathrm{B}=1$, 
and the gravitational constant $8 \pi G$ is denoted by
${\kappa}^2 \equiv 8\pi/{M_{\mathrm{Pl}}}^2$ 
with the Planck mass of $M_{\mathrm{Pl}} = G^{-1/2} = 1.2 \times 10^{19}$GeV in the following.

\section{Horndeski's theory \label{sec2}}
The action in the Horndeski's theory is given by \cite{Horndeski,Deffayet:2011gz,Kobayashi:2011nu}
\begin{equation}
S_H=\sum ^5 _{i=2} \int d ^4 x \sqrt{-g} {\cal L}_i,
\end{equation}
where
\begin{align}
{\cal L}_2 &= K(\phi , X), \\
{\cal L}_3 &= -G_3(\phi , X) \Box \phi , \\
{\cal L}_4 &= G_4(\phi , X)R + G_{4 X} \left [ ( \Box \phi )^2 - (\nabla _\mu \nabla _\nu \phi)^2 \right ], \\
{\cal L}_5 &= G_5 (\phi , X)G_{\mu \nu} \nabla ^{\mu}\nabla ^{\nu} \phi - 
\frac{G_{5X}}{6} \left [ (\Box \phi)^3 - 3 (\Box \phi) (\nabla _ \mu \nabla _ \nu \phi)^2 +2 (\nabla _ \mu \nabla _ \nu \phi)^3 \right ]. 
\end{align}
Here, $K$, $G_3$, $G_4$, and $G_5$ are generic functions of 
$\phi$ and $X=- \partial _\mu \phi \partial ^\mu \phi /2$, and the subscript $X$ means derivative with respect to $X$. 
The total action we will consider is the sum of $S_H$ and the action of matter fluid $S_\mathrm{matter}$, 
which contain baryons and cold dark matter. 
The background equations of the Universe are given by 
assuming homogeneity and isotropy of the metric. 
Substituting $\phi = \phi (t)$ and the metric $ds^2 = -N^2 (t) dt^2 + a^2 (t)dx^2$ into the action, 
and subsequently varying the action with respect to $N(t)$ gives \cite{Kobayashi:2011nu}
\begin{equation}
\rho _\mathrm{matter} + \sum ^5 _{i=2} {\cal E}_i=0 \label{FL1}, 
\end{equation}
where
\begin{align}
{\cal E}_2 &= 2XK_X - K, \\
{\cal E}_3 &=6X \dot \phi HG_{3X} -2X G_{3 \phi}, \\
{\cal E}_4 &= -6H^2 G_4 +24 H^2 X (G_{4X}+XG_{4XX})-12HX \dot \phi G_{4 \phi X}
-6H \dot \phi G_{4 \phi}, \\
{\cal E}_5 &= 2H^3 X \dot \phi (5 G_{5X}+ 2XG_{5XX}) -6 H^2 X (3G_{5 \phi} +2XG_{5 \phi X}), 
\end{align}
and $\rho _\mathrm{matter}$ is the energy density of matter. Here, 
$H=\dot a /a$ is the Hubble rate function and the dot means derivative with respect to time. 
Variation with respect to $a(t)$ yields 
\begin{equation}
p _\mathrm{matter} + \sum ^5 _{i=2} {\cal P}_i=0 \label{FL2}, 
\end{equation}
where
\begin{align}
{\cal P}_2 &= K, \\
{\cal P}_3 &= -2X \Big (  G_{3 \phi}+ \ddot \phi G_{3X} \Big), \\
{\cal P}_4 &= 2(3H^2 +2 \dot H) G_4 -4 H^2 X \bigg ( 3+ \frac{\dot X}{HX} +2 \frac{\dot H}{H^2} \bigg ) G_{4X} \nonumber \\
&-8HX \dot X G_{4XX} +2(\ddot \phi +2H \dot \phi ) G_{4 \phi }+4XG_{4 \phi \phi} + 4X(\ddot \phi -2H \dot \phi ) G_{4 \phi X}, \\
{\cal P}_5 &= -2X(2H^3 \dot \phi +2H \dot H \dot \phi +3H^2 \ddot \phi) G_{5 X}-4 H^2 X^2 \ddot \phi G_{5XX} \nonumber \\
&+4HX(\dot X -HX)G_{5 \phi X} +2 H^2 X \bigg ( 3+ 2\frac{\dot X}{HX} + 2 \frac{\dot H}{H^2} \bigg )G_{5 \phi} 
+4HX \dot \phi G_{5 \phi \phi}, 
\end{align}
and $p _\mathrm{matter}$ is the pressure of matter.
The above two Eqs.~(\ref{FL1}) and (\ref{FL2}) correspond to the Friedmann equations.  
The equation of motion of the scalar field is given by varying the action with respect to $\phi (t)$: 
\begin{equation}
\frac{1}{a^3} \frac{d}{dt} (a^3 J) = P_\phi, \label{FE}
\end{equation}
where
\begin{align}
J=& \dot \phi K_X + 6HXG_{3X} -2 \dot \phi G_{3 \phi} + 6H^2 \dot \phi (G_{4X}+2XG_{4XX})-12HXG_{4 \phi X} \nonumber \\
&+ 2H^3 X(3G_{5X} + 2XG_{5XX}) -6H^2 \dot \phi (G_{5 \phi} + XG_{5 \phi X}),
\end{align}
\begin{align}
P_{\phi} =& K_\phi -2X (G_{3 \phi \phi} + \ddot \phi G_{3 \phi X}) + 6(2H^2 + \dot H) G_{4 \phi} +6H(\dot X +2HX) G_{4 \phi X} \nonumber \\
&- 6H^2 X G_{5 \phi \phi} + 2H^3 X \dot \phi G_{5 \phi X}.
\end{align}
Equations (\ref{FL1}), (\ref{FL2}), and (\ref{FE}) control the background evolution of the Universe. 
In the same manner as the quintessence model, Eqs.~(\ref{FL2}) and (\ref{FE}) are equivalent when Eq.~(\ref{FL1}) holds. 
Equations (\ref{FL1}) and (\ref{FL2}) can be rewritten in the well-known form
\begin{align}
3H^2 = \kappa ^2 (\rho _\mathrm{matter} + \rho _\phi), \\ 
-3H^2 -2 \dot H = \kappa ^2 (p_\mathrm{matter} + p_\phi),
\end{align}
if we define $\rho _{\phi}$ and $p _\phi$ as 
\begin{equation}
\rho _{\phi} \equiv \sum ^5 _{i=2} {\cal E}_i + \frac{3H^2}{\kappa ^2}, \qquad 
p_{\phi} \equiv \sum ^5 _{i=2} {\cal P}_i - \frac{1}{\kappa ^2} (3H^2 + 2 \dot H). \label{rhop}
\end{equation}
We will use Eq.~(\ref{rhop}) as the definitions of effective energy density and effective pressure. 
In the following, we will only consider the matter dominant era onwards and treat the pressure of matter as $p_\mathrm{matter} =0$. 

While, the recent observation of gravitational wave GW170817 \cite{TheLIGOScientific:2017qsa}
and its electromagnegic counterparts \cite{Monitor:2017mdv,GBM:2017lvd,Coulter:2017wya} showed that the speed of 
gravitational waves should satisfy 
\begin{equation}
\vert c_T^2 -1 \vert \lesssim 10^{-15} 
\end{equation}
in the relatively recent Universe. 
This bound means that the speed of gravitational waves should be almost the same as that of an electromagnetic wave not only 
around stellar objects but also in the void region.  
In Horndeski's theory, the sound speed for tensor perturbations is expressed as \cite{Kobayashi:2011nu}
\begin{equation}
c_T ^2 = \frac{G_4 -XG_{5 \phi}-XG_{5X} \ddot \phi}{G_4 -2XG_{4X}-X(G_{5X}\dot \phi H - G_{5 \phi})}. 
\label{ct}
\end{equation}
Equation (\ref{ct}) shows that the sound speed for tensor perturbations is independent from 
the functions $K(\phi ,X)$, $G_3 (\phi , X)$, and matter components. 
If the terms $XG_{5 \phi}$, $XG_{4X}$, $\cdots$  are relevant for the evolution of the Universe, then 
they should be the same order as $G_4$ as seen from Eqs.~(\ref{FL1}) and (\ref{FL2}). 
In this case, $c_T^2$ deviates from $1$ except for some special cases. 
Therefore, it is natural to think that the terms proportional to $G_{4X}$, $G_{5 \phi}$, and $G_{5 X}$ are 
not relevant for the current accelerated expansion of the Universe.  
In the following, we treat $G_4(\phi , X)$ and $G_5(\phi, X)$ as $G_4(\phi)$, 
which is a generic function only depends on the field $\phi$, 
and $G_5(\phi , X)=0$. Here,  
$G_5 (\phi , X)$ is not expressed as constant but as $0$ because constant $G_5$ does not contribute to Eqs.~(\ref{FL1}), (\ref{FL2}), and (\ref{FE}).  
Further discussions for the constraints from gravitational wave detection GW170817 for Horndeski's theory are given in Refs.~\cite{Creminelli:2017sry,Sakstein:2017xjx,Ezquiaga:2017ekz,Baker:2017hug,Arai:2017hxj,Langlois:2017dyl,Lombriser:2015sxa,Lombriser:2016yzn}. 
Then, effective energy density and effective pressure are explicitly written by
\begin{align}
\rho _{\phi} &= \dot \phi ^2 K_X -K +3 \dot \phi ^3 H G_{3X} - \dot \phi ^2 G_{3 \phi} + 3H^2 \left ( \frac{1}{\kappa ^2} -2G_4 \right ) -6H \dot \phi G_{4 \phi}, \\
p_{\phi} &= K- \dot \phi ^2 (G_{3 \phi} + \ddot \phi G_{3X}) -(3H^2 + 2 \dot H) \left ( \frac{1}{\kappa ^2} -2G_{4} \right ) +2( \ddot \phi +2H \dot \phi ) G_{4 \phi} +2 \dot \phi ^2 G_{4 \phi \phi}. 
\end{align}
Equation of motion of the scalar field Eq.~(\ref{FE}) is rewritten as
\begin{equation}
\frac{d}{a^3 dt} \left [ a^3 ( \dot \phi K_X + 3H \dot \phi ^2 G_{3X} -2 \dot \phi G_{3 \phi} ) \right ] -K_\phi + \dot \phi ^2 (G_{3 \phi \phi} + \ddot \phi G_{3 \phi X}) 
- 6(2H^2 + \dot H) G_{4 \phi} = 0. 
\end{equation}

The perturbation equations are given by substituting flat FLRW metric with Newtonian gauge metric perturbation $ds^2 = -(1+2 \Psi (t,x)) dt^2 + a^2(t) (1+2 \Phi (t,x)) \sum _{i,j=1} ^3 \delta _{ij} dx^i dx^j$  
into the linear order Einstein equation $(\delta G_4 G_\mu ^{\; \; \nu} + G_4 \delta G_\mu ^{\; \; \nu} + \cdots - \delta T_{\quad \mu} ^{(m) \nu}/2 =0)$ as follows: 
\begin{align}
A_1 \dot \Phi + A_2 \dot {\delta \phi}  + A_3 \frac{k^2}{a^2} \Phi 
+A_4 \Psi + \left (A_6 \frac{k^2}{a^2}- \mu \right ) \delta \phi - \rho _\mathrm{m} \delta =0, \label{le00} \\
B_1 \ddot \Phi + B_2 \ddot {\delta \phi} +B_3 \dot \Phi + B_4 \dot {\delta \phi} + B_5 \dot \Psi 
+B_6 \frac{k^2}{a^2} \Phi \nonumber \\
+ \left ( B_7 \frac{k^2}{a^2} +3 \nu \right ) \delta \phi 
+ \left ( B_8 \frac{k^2}{a^2} + B_9 \right ) \Psi =0, \label{leii} \\
C_1 \dot \Phi + C_2 \dot {\delta \phi} + C_3 \Psi + C_4 \delta \phi - \rho _\mathrm{m} \delta u =0, \label{le0i} \\
G_4 (\Psi + \Phi ) + G_{4 \phi} \delta \phi =0, \label{leij} 
\end{align}
where the coefficients $A_1$, $A_2$, $\cdots$, $B_1$, $\cdots$, $C_1$, $\cdots$, and $\nu$ are the functions composed from $H$, $K(\phi , X)$, $G_3 (\phi , X)$, and $G_4 (\phi)$, 
and $k$ expresses the wave number which appears from the derivative with respect to 
the spacial coordinates ($k^2 = -\partial _j \partial _j$) by the Fourier transformation. 
The concrete expressions of the coefficients are given in Appendix \ref{app1}. Equations 
(\ref{le00}), (\ref{leii}), (\ref{le0i}), and (\ref{leij}) correspond to $(0,0)$, $(i,i)$, $(0,i)$, and $(i,j)$, $i \neq j$ components of the Einstein equations, respectively, 
and are consistent with the equations derived in Ref.~\cite{DeFelice:2011hq}. 
We have adopted $\delta \rho _\mathrm{m} = \rho _\mathrm{m} \delta $, $\delta p_\mathrm{m} =0$, 
and $\delta u_i = \partial _i \delta u$ as the descriptions of $\delta T_{\quad \mu} ^{(m) \nu}= 
\delta [p_\mathrm{m} \delta_{\mu}^{\; \nu} + (\rho _\mathrm{m} +p_\mathrm{m})u_\mu u^\nu]$ 
because we are now interested in the scalar perturbations from nonrelativistic matter fluids. 
While, the linearized scalar field equation and the linearized equations of continuity of fluid matter are expressed by 
\begin{align}
D_1 \ddot \Phi + D_2 \ddot {\delta \phi} +D_3 \dot \Phi + D_4 \dot {\delta \phi} + D_5 \dot \Psi 
+ \left ( D_7 \frac{k^2}{a^2} +D_8 \right ) \Phi \nonumber \\
+ \left ( D_9 \frac{k^2}{a^2} -M^2 \right ) \delta \phi 
+ \left ( D_{10} \frac{k^2}{a^2} + D_{11} \right ) \Psi =0, \label{lsf} \\
\dot \delta + 3 \dot \Phi  - \frac{k^2}{a^2} \delta u =0, \label{lt0} \\
\dot { \delta u} + \Psi =0, \label{lti} 
\end{align}
where the coefficients $D_1$, $D_2$, $\cdots$, and $M^2$ are the functions of background quantities $H(t)$, $K(\phi (t) , X(t))$, $G_3 (\phi (t) , X(t))$, and $G_4 (\phi (t))$ 
and their definitions are given in Appendix \ref{app1}. 
Usually, we treat the coefficients $H D_1$, $H^2 D_2$, $\cdots$, and $M^2$ as the same order, for example, $H^2 D_1 \sim H D_3 \sim H^2 D_7 \sim D_8$, 
because the coefficients are related with the background evolution of the Universe. 
Therefore, we can ignore the term proportional to $D_8$ in the subhorizon limit $k^2/(a^2 H^2) \gg 1$. 
However, in some special cases, e.g. in the case of $f(R)$ gravity, function $M^2$ should be treated as much important as $k^2 D_9 /a^2$, because 
the observational constraints on the function $f(R)$; $|R_0f''(R_0) | \ll 1$, enhances the value of $M^2$ \cite{DeFelice:2011hq}. 
\section{Evolution equation of the matter density contrast \label{sec3}}
\subsection{Quasi-static solutions}
Before considering the general solutions of Eqs.~(\ref{le00}) -- (\ref{lti}), let us investigate quasi-static solutions of 
the system, because quasi-static solutions are simple and easy to treat.
Here, quasi-static means that the perturbation quantities, e.g. $\Phi$, $\Psi$, and $\delta \phi$, vary in the same way as the Hubble rate, namely, 
$ \dot \Phi \sim H \Phi$, $\dot \Psi \sim H \Psi$, and $\dot {\delta \phi } \sim H \delta \phi$. 
One can easily imagine that there is such a solution, because the coefficients 
$A_1$, $A_2$, $\cdots$, $B_1$, $\cdots$, $C_1$, $\cdots$, $D_1$, $\cdots$, and so on are described by 
background quantities which vary in the same way as the Hubble rate. Whereas, a hierarchy between the coefficients caused by the factor $k^2/(a^2H^2)$ 
implies the existence of fast varying solutions, which is oscillating mode that we will consider in the next subsection. 
Under the quasi-static conditions, Eqs.~(\ref{le00}) -- (\ref{lti}) are rewritten as 
\begin{align}
A_3 \frac{k^2}{a^2} \Phi 
+A_4 \Psi + A_6 \frac{k^2}{a^2} \delta \phi - \rho _\mathrm{m} \delta \simeq 0, \label{le00qs} \\
B_6 \frac{k^2}{a^2} \Phi 
+ B_7 \frac{k^2}{a^2}  \delta \phi 
+ B_8 \frac{k^2}{a^2} \Psi \simeq 0, \label{leiiqs} \\
C_1 \dot \Phi + C_2 \dot {\delta \phi} + C_3 \Psi + C_4 \delta \phi - \rho _\mathrm{m} \delta u =0, \label{le0iqs} \\
G_4 (\Psi + \Phi ) + G_{4 \phi} \delta \phi =0, \label{leijqs} \\
D_7 \frac{k^2}{a^2}  \Phi
+ \left ( D_9 \frac{k^2}{a^2} -M^2 \right ) \delta \phi 
+ D_{10} \frac{k^2}{a^2} \Psi \simeq 0, \label{lsf	qs} \\
\dot \delta + 3 \dot \Phi  - \frac{k^2}{a^2} \delta u =0, \label{lt0qs} \\
\dot { \delta u} + \Psi =0. \label{ltiqs} 
\end{align}
Here, we have kept the term $M^2$ by taking into account the case of $f(R)$ gravity. The reason why $M^2$ should be kept is mentioned in the last part of the previous section. 
In this paper, we will not use gauge-invariant density contrast defined by 
$\delta _m \equiv \delta -3H \delta u$ because the relation $| \delta | \gg H | \delta u |$ holds not only in 
quasi-static solutions but also in oscillating solutions as shown later. 
Therefore, there is no difference in the prediction of the growth rate function between $\delta _m$ and $\delta$ at the leading order. 

The so-called effective gravitational constant $G_\mathrm{eff} $ defined by 
$k^2 \Psi /a^2 = -4 \pi G_\mathrm{eff} \rho _\mathrm{m} \delta$ is given by eliminating the terms 
proportional to $\Phi$ or $\delta \phi$ from Eq.~(\ref{le00qs}). 
The explicit expression of $G_\mathrm{eff} $ is
\begin{align}
 G_\mathrm{eff}=\frac{1}{16 \pi G_4 \beta} \bigg \{ G_4 [K_X-2G_{3 \phi} + \ddot \phi (2 G_{3X} + \dot \phi ^2 G_{3XX}) + \dot \phi ^2 G_{3 \phi X} + 4H \dot \phi G_{3X} ] 
 \nonumber \\
 +4 G_{4 \phi}^2  + \frac{a^2}{k^2} G_4 M^2 \bigg \}, \label{geff} \\
 \beta \equiv G_4 [K_X -2G_{3 \phi} + \ddot \phi (2G_{3X} + \dot \phi ^2 G_{3XX})+ \dot \phi ^2 G_{3 \phi X} +4H \dot \phi G_{3X}] 
 - \frac{1}{4} \dot \phi ^4 G_{3X}^2  \nonumber \\
 - \dot \phi ^2 G_{3X} G_{4 \phi} +3 G_{4 \phi} ^2 + \frac{a^2}{k^2} G_4 M^2. 
\end{align}
Equation (\ref{geff}) shows that if there is an $X$ dependence in $G_{3}$ function or a $\phi$ dependence in $G_4$ function, then $G_\mathrm{eff}$ can deviate from 
the Newtonian constant even if we ignore the term proportional to $M^2$. 
The evolution equation of the matter density contrast, which is given by combining Eqs.~(\ref{lt0qs}) and (\ref{ltiqs}), is expressed as 
\begin{equation}
\ddot \delta + 2H \dot \delta -4 \pi G_\mathrm{eff} \rho _\mathrm{m} \delta \simeq 0. 
\label{eqqs}
\end{equation}
Equation (\ref{eqqs}) shows that the modification of the evolution equation is only the change of the Newtonian gravitational constant. 
While, $\alpha$ parameter defined by 
$\Psi + \Phi = \alpha \Phi$, which describes the modification of the lensing effect, is written as 
\begin{align}
\alpha = -G_{4 \phi }(2G_{4 \phi}+ \dot \phi ^2 G_{3X}) \bigg / \bigg \{ G_4 [K_X-2G_{3 \phi}+ \ddot \phi (2G_{3X}+ \dot \phi ^2 G_{3XX})+ \dot \phi ^2 G_{3 \phi X} 
\nonumber \\
+4H \dot \phi G_{3X}]+ G_{4 \phi}(- \dot \phi ^2G_{3X}+ 2G_{4 \phi})+ \frac{a^2}{k^2} G_4 M^2 \bigg \} .
\end{align}
Unlike in the case of $G_\mathrm{eff}$, $\alpha$ can deviates from $\alpha = 0$ (GR) only if $G_{4 \phi} \neq 0$.  
Therefore, for the case $G_{3 X} \neq 0$ and $G_{4 \phi} = 0$, there is a modification for the growth rate function but 
is no change in the lensing effect. 
In this subsection, we have respected the possibility that $M^2$ can be comparable with $k^2/a^2$, 
however, for simplicity, we will assume $a^2 M^2  /k^2 \ll 1$ in the following. Oscillating solutions in $f(R)$ modified gravity are investigated in Ref.~\cite{Matsumoto:2013sba}. 
\subsection{$k$-essence model}
$k$-essence model \cite{ArmendarizPicon:1999rj,Garriga:1999vw,ArmendarizPicon:2000ah} corresponds to the case $G_3(\phi , X)=0$ and $G_4(\phi) = 1/(2 \kappa ^2)$ in Horndeski's theory, so there is only one generic function $K(\phi ,X)$.  
In $k$-essence model, it is shown that the general solution of the matter density contrast is given in the following form \cite{Bamba:2011ih}
\begin{align}
\delta _{tot} (N)= \delta _{qs} (N) + \delta _o (N), \label{deltatot} \\
\frac{d ^2 \delta _{qs}}{dN^2} (N) + \left ( \frac{1}{2} -\frac{3}{2} w_\mathrm{eff} (N) \ \right ) \frac{d \delta _{qs}}{dN} (N) 
-\frac{3}{2} \Omega _\mathrm{m} (N) \delta _{qs} (N) =0, \label{kqs} \\
\delta _o (N) = c_1 \dot \phi (N) [4K_X (N)+ \dot \phi ^2 (N) K_{XX}(N)]K_X^{- \frac{3}{4}}(N) [K_X(N) + \dot \phi ^2(N) K_{XX}(N)]^\frac{1}{4} \nonumber \\
\times \sin \left [ \int ^N dN' \frac{c_\phi k}{aH} + c_2  \right ] , \label{ocisk}
\end{align}
where $c_1$ and $c_2$ are arbitrary constants, $N$, $w_\mathrm{eff}$, $\Omega_\mathrm{m}$, and $c_\phi$ are defined by 
$N \equiv \ln a$, $w_\mathrm{eff} \equiv -1-2 \dot H/(3H^2)$, $\Omega_\mathrm{m} \equiv \kappa ^2 \rho _\mathrm{m} /(3H^2)$, 
and $c_\phi ^2 \equiv K_X/(K_X+\dot \phi ^2 K_{XX})$, respectively. 
One may think the oscillation term can be ignored by assuming $c_1 =0$.  However, there is no guiding principle which supports 
$| c_1 | \ll 1$. In other words, assuming $c_1 =0$ is realized only if we introduce fine-tunings of the initial conditions. 
Therefore, if we choose $c_1 =0$ it will introduce 
another fine-tuning problem of dark energy. The 
explicit expression of $c_1$ will be given in Eq.~(\ref{cdeltak}).  
In the paper \cite{Bamba:2011ih}, the fourth order evolution equation of the matter density contrast is derived without using approximations, 
that's why, not only well known quasi-static equation (\ref{kqs}) but also oscillating solutions (\ref{ocisk}) appear. 
The reason why there is an oscillating mode is that there are extra degrees of freedom, which correspond to 
the perturbation of dynamical scalar field $\delta \phi$ and its derivatives with non-zero sound speed, compared to General Relativity. 
Evaluation of the oscillating mode is much more difficult than that of the quasi-static mode, because 
not only the leading terms of the linearized equations but also the sub-leading terms should be taken into account 
in order to obtain the effective growth rate of the oscillating solutions. 
One way to evaluate the behavior of the oscillating solution is to derive the fourth order evolution equation of the matter density contrast 
as shown in \cite{Bamba:2011ih}. 
This way is correct, but it costs so much time. 
We will construct the other way to derive Eq.~(\ref{ocisk}) in the following, and will apply it to whole Horndeski's theory in the next subsection. 

The oscillating mode of the matter density contrast is generated by the mixing between $\delta$ and $\delta \phi$ in the linearized equations, 
and the oscillating interval is determined by the sound speed.  
By taking into account $k/(aH) \gg 1$, we can write the solution of the oscillating mode in the following form: 
\begin{equation}
\delta _o (N) = C_\delta (N)  \left \{ \cos \left [ \int ^N dN' \frac{c_s k}{aH} + \omega  \right ] + O \left ( \frac{aH}{k} \right ) \right \}, 
\label{odk}
\end{equation}
where $\omega$ is an arbitrary constant and $C_\delta (N)$ is an arbitrary function which  
satisfies $| dC_\delta /dN | \lesssim |C_\delta |$, namely, 
\begin{equation}
\frac{d \delta _o}{ dN}  \sim C_\delta (N) \frac{d}{dN} \cos \left [ \int ^N dN' \frac{c_s k}{aH} + \omega  \right ]. 
\end{equation}
Equation (\ref{odk}) is, in fact, an WKB approximated solution. 
The WKB approximation is valid only if $|c_s k/(aH)| \gg 1$. In a typical dark energy model ``quintessence model", the sound speed $c_s$ is always $1$, therefore, 
the condition $|c_s k/(aH)| \gg 1$ becomes same as small-scale approximation $|k/(aH)| \gg 1$. 
The deviation from the exact solution is given by the order of $|aH/k|$. 
If we consider the model which has a little $|c_s|$, then the WKB approximation breaks down. 
However, the fine-tunings of the model parameters are necessary to realize $c_s^2 \ll 1$. The purpose of this paper is to investigate general perturbation behavior in Horndeski's theory, therefore, investigating the case $c_s^2 \ll 1$ is beyond the scope of this paper.
In the same way, we will describe the oscillating solution of $\delta \phi$ and $\Phi$ as
\begin{align}
\delta \phi _o (N) = C_{\delta \phi}(N) \phi \left \{ \sin \left [ \int ^N dN' \frac{c_s k}{aH} + \omega  \right ] + O \left ( \frac{aH}{k} \right ) \right \}, \label{dphiok} \\
\Phi _o (N) = C_\Phi (N)  \left \{ \cos \left [ \int ^N dN' \frac{c_s k}{aH} + \omega  \right ] + O \left ( \frac{aH}{k} \right ) \right \}. \label{phiok}
\end{align}
Equations (\ref{odk})--(\ref{phiok}) give the relations $\ddot \delta _o \sim -c_s^2 k^2 \delta _o /a^2$, $\ddot {\delta \phi} _o \sim -c_s^2 k^2 \delta \phi _o /a^2$, 
and $\ddot \Phi _o \sim -c_s^2 k^2 \Phi _o /a^2$. 
The reason why $\sin$ and $\cos$ are assigned for $\delta \phi$ and $\Phi$, respectively, will be shown in the following considerations. 
In the case of $k$-essence model, Eq.~(\ref{leij}) gives $\Psi = -\Phi$, therefore, we will write $\Psi$ as $- \Phi$ in this subsection.

Equation of motion of the scalar field (\ref{lsf}) yields 
\begin{align}
\delta \ddot \phi + \left ( 3H + \frac{d}{dt} \ln |K_X + \dot \phi ^2 K_{XX} | \right ) \delta \dot \phi 
+ \frac{c_\phi ^2 k^2}{a^2} \delta \phi + \frac{4 \dot \phi K_X + \dot \phi ^3 K_{XX}}{K_X + \dot \phi ^2 K_{XX}} \dot \Phi \simeq 0, \label{klsf} 
\end{align}
where $c_\phi ^2 = K_X/(K_X + \dot \phi ^2 K_{XX})$, which is the sound speed squared in $k$-essence model. Here, the symbol $\simeq$ means that $O(H^2 \delta \phi)$ terms have been ignored. 
While, Eq.~(\ref{leii}) gives 
\begin{equation}
\ddot \Phi + 4H \dot \Phi - \frac{\kappa ^2}{2} K_X \dot \phi \delta \dot \phi \simeq 0. \label{Pdpr0}
\end{equation}
At the leading order, we have 
\begin{equation}
\vert \ddot \Phi \vert \simeq  \left \vert  \frac{\kappa ^2}{2} K_X \dot \phi \delta \dot \phi \right \vert .  \label{Pdpr1}
\end{equation}
By taking Eqs.~(\ref{dphiok}) and (\ref{phiok}) into account, we have
\begin{equation}
\vert \dot \Phi  \vert \simeq  \left \vert  \frac{\kappa ^2}{2} K_X \dot \phi \delta \phi \right \vert . \label{Pdpr2}
\end{equation}
The relations (\ref{Pdpr1}) and (\ref{Pdpr2}) show that if $\delta \phi$ is given by sinusoidal function, then $\Phi$ should 
be described by cosine function, and imply that the last term in Eq.~(\ref{klsf}) is $O(H^2 \delta \phi)$. 
Then we can solve Eq.~(\ref{klsf}) by using the WKB approximation. The coefficient $C_{\delta \phi} (N)$ in Eq.~(\ref{dphiok})
is expressed as
\begin{equation}
C_{\delta \phi} (N) = m_{\delta \phi} \left [ a \phi \sqrt{c_\phi (K_X + \dot \phi ^2 K_{XX})} \right ] ^{-1}, \label{c1}
\end{equation}
where $m_{\delta \phi}$ is an arbitrary constant. 
While, the relation (\ref{Pdpr1}) or (\ref{Pdpr2}) gives 
\begin{equation}
C_{\Phi} (N) = - \frac{a}{2 c_\phi k} \kappa ^2 \phi \dot \phi K_X C_{\delta \phi} (N). \label{cphik}
\end{equation}
Note that the expression (\ref{cphik}) is determined only by the leading terms in Eq.~(\ref{Pdpr0}), 
and whole equation (\ref{Pdpr0}) is completed by taking $O ( aH/k )$ terms in Eq.~(\ref{phiok}) into account. 
To obtain the relation between $C_{\Phi}$ and $C_\delta$, we can use Eqs.~(\ref{lt0}) and (\ref{lti}). 
 Eliminating $\delta u$ from Eqs.~(\ref{lt0}) and (\ref{lti}) yields
\begin{equation}
\ddot \delta + 2H \dot \delta + 3 \ddot \Phi +6H \dot \Phi - \frac{k^2}{a^2} \Phi =0. \label{mdpk}
\end{equation}
At the leading order in Eq.~(\ref{mdpk}), we have
\begin{equation}
\frac{d}{dt} (a^2 \dot \delta ) \simeq k^2 (1+3c_\phi ^2) \Phi
\end{equation}
\begin{equation}
\delta \simeq - \frac{1+3c_\phi ^2}{c_\phi ^2} \Phi .
\end{equation}
Therefore, 
\begin{align}
C_\delta (N) &= - \frac{1+3c_\phi ^2}{c_\phi ^2} C_{\Phi} (N) \nonumber \\
&= \frac{m_{\delta \phi} \kappa ^2}{2 k} \dot \phi (4K_X + \dot \phi ^2 K_{XX}) K_X^{-\frac{3}{4}}(K_X + \dot \phi ^2 K_{XX})^\frac{1}{4}. 
\label{cdeltak}
\end{align}
The expression of $C_\delta (N)$ (\ref{cdeltak}) is consistent with Eq.~(\ref{ocisk}). 
Note that Eqs.~(\ref{cphik}) and (\ref{cdeltak}) show that $|C _\delta | \sim |C_\Phi | \sim | aH C_{\delta \phi}/k |$ 
if $c_\phi \neq 0$. Therefore, Eq.~(\ref{lti}) shows that $|H\delta u / \delta| \sim aH /k \ll 1$. The effective growth factor of the matter density perturbation can be defined as 
\begin{equation}
f_\mathrm{eff} = \frac{d}{dN} \ln | C_{\delta } (N)| \label{feff}
\end{equation}
by comparing it with the usual growth factor $f=d \ln | \delta _{qs}(N)|/dN$. 
The usual growth factor express the growing speed of the matter density perturbation, while, the effective growth factor express the 
growing speed of the amplitude of the matter density perturbation in oscillating mode. 
The effective growth factor is only expressed by background quantities $\dot \phi (t)$ and $\ddot \phi (t)$, and $H(t)$ as seen from Eqs.~(\ref{cdeltak}) and (\ref{feff}),  
therefore, the behavior of the effective growth factor is independent from the initial conditions of the perturbation quantities. 
Interestingly, it is also seen that the effective growth factor does not explicitly depend on the potential form of the scalar field. 
If $f_\mathrm{eff}$ is same order as the usual growth factor $f$, then the amplitude $C_\delta (N)$ is same order as the absolute value of the density contrast of the quasi-static mode 
if there is no hierarchy in the initial conditions of them. 
In this case, we cannot neglect the term $\delta _o$ in Eq.~(\ref{deltatot}). In observations, the power spectrum of the density contrast of galaxies, which is approximately proportional to the 
power spectrum of the density contrast of matter, is measured. 
Therefore, comparing the evolution of $f_\mathrm{eff}$ with that of $f$ is crucial. 
Even if it is shown that $f_\mathrm{eff}$ is much less than $f$, this comparison gives us a confirmation that the quasi-static approximation is good enough. 

The features of the procedure above are the following; we have assumed the WKB approximated solution because of non-zero sound speed, 
not only the leading terms but also the sub-leading terms are evaluated, 
the coefficient $C_\delta (N)$ is uniquely determined by considering the large/small relation between the perturbation variables. 
We will apply this procedure to the other case in the next subsection. 

\subsection{The case $G_{3X} \neq 0$ or $G_{4 \phi} \neq 0$}
The case $G_{3X}=G_{4 \phi} =0$ is equivalent to the case of $k$-essence, so we will investigate the case 
$G_{3X} \neq 0$ or $G_{4 \phi} \neq 0$ in this subsection. 
The relations between the perturbation variables are quite different from those in the case of $k$-essence model. 
As seen in Eq.~(\ref{leii}), 
function $B_2$ does not vanish and $\Phi$ becomes same order as $\delta \phi$, 
because $B_2$ is expressed as $-3 \dot \phi ^2 G_{3X}+6G_{4 \phi}$.
Therefore, we cannot ignore the terms proportional to $\Psi$, $\Phi$, and derivatives of them in Eq.~(\ref{lti}), 
and cannot solve Eq.~(\ref{leii}) or Eq.~(\ref{lti}), independently. 
In this case, we should take into account the 
sub-leading terms in Eqs.~(\ref{dphiok}) and (\ref{phiok}), because there is a possibility that 
the sub-leading terms change the expressions of $C_{\delta \phi}$, $C_{\Phi}$, and $C_{\delta}$. 

In the following, we will use Eq.~(\ref{leij}) to erace $\Psi$, and express $\Phi$ and $\delta \phi$ as 
\begin{align}
\Phi = C_{\Phi 1} \sin \left [ \int ^N dN' \frac{c_s k}{aH} + \omega \right ] + C_{\Phi 2} \cos \left [ \int ^N dN' \frac{c_s k}{aH} + \omega \right ] , \\
\delta \phi = C_{\delta \phi 1} \sin \left [ \int ^N dN' \frac{c_s k}{aH} + \omega \right ] + C_{\delta \phi 2} \cos \left [ \int ^N dN' \frac{c_s k}{aH} + \omega \right ] .
\end{align}
where $C_{\Phi 2} / C_{\Phi 1} \sim O(aH/k)$ and $C_{\delta \phi 2} / C_{\delta \phi 1} \sim O(aH/k)$ have been assumed. 
Equations (\ref{leii}) and (\ref{lti}) give 
\begin{align}
B_{f2} \ddot \Phi + B_{f1} \dot \Phi +B_{d2} \delta \ddot \phi + B_{d1} \delta \dot \phi \simeq 0, \label{eb} \\
D_{f2} \ddot \Phi + D_{f1} \dot \Phi + D_{f0} \frac{k^2}{a^2} \Phi + D_{d2} \delta \ddot \phi + D_{d1} \delta \dot \phi + D_{d0} \frac{k^2}{a^2} \delta  \phi \simeq 0, \label{ed}
\end{align}
where $B_{f2}=B_1$, $B_{f1}=B_3-B_5$, $B_{d2}=B_2$, $B_{d1}=B_4-B_5 \cdot G_{4 \phi}/G_4$, 
$D_{f2}=D_1$, $D_{f1}=D_3-D_5$, $D_{f0}=D_7-D_{10}$, $D_{d2}=D_2$, $D_{d1} = D_4-D_5 \cdot G_{4 \phi}/G_4$, and 
$D_{d0}=D_9-D_{10} \cdot G_{4 \phi}/G_4$. 
Here, the symbol $\simeq$ means that $O(H^2 \Phi)$ terms and $O(H^2 \delta \phi)$ terms have been ignored. 
The leading order terms, which are the terms proportional to $k^2/a^2$, in Eq.~(\ref{eb}) yield the ratio between $C_{\Phi 1}$ and $C_{\delta \phi 1}$ as 
\begin{equation}
\gamma \equiv \frac{C_{\Phi 1}}{C_{\delta \phi 1}} = - \frac{B_{d2}}{B_{f2}}= - \frac{2G_{4 \phi} - \dot \phi ^2 G_{3X}}{4G_4}. 
\end{equation}
On the other hand, cancellation of the leading terms in Eq.~(\ref{ed}) induces the following expression of sound speed: 
\begin{align}
c_s^2 = & \; \frac{D_{f0} \gamma + D_{d0}}{D_{f2} \gamma + D_{d2}} \nonumber \\
=& \; \frac{1}{\xi} \bigg [ G_4 (K_X-2G_{3 \phi} +2 \ddot \phi G_{3X} + \dot \phi ^2 G_{3 \phi X} + \dot \phi ^2 \ddot \phi G_{3XX} +4 H \dot \phi G_{3X}) \nonumber \\
&+3G_{4 \phi}^2 - \dot \phi ^2 G_{3X} G_{4 \phi} - \frac{1}{4} \dot \phi ^4 G_{3X}^2 \bigg ], \label{cs} \\
\xi \equiv & \; G_4 \bigg [  K_X + \dot \phi ^2 K_{XX} -2G_{3 \phi} - \dot \phi ^2 G_{3 \phi X} + 3H \dot \phi (2G_{3X}+ \dot \phi ^2 G_{3XX}) \bigg ] \nonumber \\
&+3 \left ( G_{4 \phi} - \frac{1}{2} G_{3X} \dot \phi ^2  \right )^2. \label{csd}
\end{align}
Note that $\xi >0$ is imposed by no-ghost condition and $c_s^2$ should not be negative to keep the validity of the perturbation theory. 
While, at the sub-leading order, Eqs.~(\ref{eb}) and (\ref{ed}) are written as 
\begin{align}
B_{f2} (C_{\Phi 2} - \gamma C_{\delta \phi 2}) \frac{c_s k}{a} =& \; (2B_{f2} \dot \gamma + B_{f1} \gamma + B_{d1}) C_{\delta \phi 1}, \label{ebs} \\
C_{\delta \phi 1} \frac{1}{dt} \ln \left | \frac{c_s}{a}C_{\delta \phi 1}^{\; 2} \right | =& - \frac{2D_{f2} \dot \gamma + D_{f1} \gamma +D_{d1}}{D_{f2} \gamma + D_{d2}}C_{\delta \phi 1}
\nonumber \\ &+ \frac{(C_{\Phi 2}- \gamma C_{\delta \phi 2})(D_{f2}D_{d0}-D_{f0}D_{d2})}{(D_{f2}\gamma + D_{d2})^2} \frac{k}{c_s a} \label{eds} .
\end{align}
Eliminating the terms $C_{\Phi 2} - \gamma C_{\delta \phi 2}$ by combining Eqs.~(\ref{ebs}) and (\ref{eds}) yields 
\begin{align}
\frac{1}{dt} \ln \left | \frac{c_s}{a}C_{\delta \phi 1}^{\; 2} \right | =& - \frac{2D_{f2} \dot \gamma + D_{f1} \gamma +D_{d1}}{D_{f2} \gamma + D_{d2}} 
+ \frac{(D_{f2}D_{d0}-D_{f0}D_{d2})(2B_{f2} \dot \gamma +B_{f1} \gamma +B_{d1})}{B_{f2}(D_{f0}\gamma + D_{d0})(D_{f2}\gamma + D_{d2})} . \label{cdeltaf}
\end{align}
Equation (\ref{cdeltaf}) shows that the expression of $C_{\delta \phi 1}$ does not depend on the forms of $C_{\delta \phi 2}$ and $C_{\Phi 2}$ 
and is uniquely determined except for the integration constant. 
Substituting the expressions of $B_{f1}$, $B_{f2}$, $\cdots$ and $D_{f1}$, $D_{f2}$, $\cdots$ into Eq.~(\ref{cdeltaf}) gives 
\begin{equation}
C_{\delta \phi 1} ^{\; 2} = \;  \frac{m_{\delta \phi 1}^2 G_4}{4 c_s a^2 \xi }, \label{cdeltaphi1}
\end{equation}
where $m_{\delta \phi 1}^2$ is an integration constant. Whereas, we can obtain the relation between $C_{\delta \phi 1}$ and $C_{\delta}$ from Eqs.~ (\ref{lt0}) and (\ref{lti}) as 
follows: 
\begin{align}
\frac{1}{a^2}\frac{d}{dt}(a^2 \dot \delta) =& -3 \ddot \Phi -6H \dot \Phi - \frac{k^2}{a^2} \Psi \nonumber \\
\simeq & \; \frac{k^2}{a^2} \left [ (1+3 c_s ^2) \Phi + \frac{G_{4 \phi}}{G_4} \delta \phi \right ] ,  
\end{align}

therefore, 

\begin{align}
 C_\delta = - \frac{1}{c_s ^2} \left [ (1+3 c_s ^2) \gamma + \frac{G_{4 \phi}}{G_4}  \right ] C_{\delta \phi 1} , \label{cdeltacphi}
\end{align}

where we have assumed the form of $\delta _o$ as 
\begin{align}
\delta _o (N) = C_\delta (N)  \left \{ \sin \left [ \int ^N dN' \frac{c_s k}{aH} + \omega  \right ] + O \left ( \frac{aH}{k} \right ) \right \} .
\end{align}
By substituting Eq.~(\ref{cdeltaphi1}) into Eq.~(\ref{cdeltacphi}), we finally obtain
\begin{equation}
C_{\delta} = \mp \frac{m_{\delta \phi 1}}{8a \sqrt{\xi c_s G_4}} \left [ c_s^{-2} ( \dot \phi ^2 G_{3X}+2G_{4 \phi}) +3(\dot \phi ^2 G_{3X} -2 G_{4 \phi})  \right ] . 
\label{cdeltagen}
\end{equation}
Note that the expression (\ref{cdeltagen}) is only valid for the case $G_{4 \phi} \neq 0$ or $G_{3 X} \neq 0$, 
because we treat $\Phi$ same order as $\delta \phi$ in the very begging of this subsection by taking $G_{4 \phi} \neq 0$ or $G_{3 X} \neq 0$ into account. 
This relation between $\Phi$ and $\delta \phi$ is seriously changed from that in the case $G_{4 \phi}= G_{3 X}=0$. 
It means that the sub-leading terms in the case $G_{4 \phi}= G_{3 X}=0$ can be the leading terms in the case 
$G_{4 \phi} \neq 0$ or $G_{3X} \neq 0$.  Therefore, we cannot connect the expression (\ref{cdeltagen}) with (\ref{cdeltak}).
Expression of the effective growth rate function is obtained by using Eqs.~(\ref{feff}) and (\ref{cdeltagen}). 
The effective growth rate function is expressed by background quantities $\phi (t)$, $\dot \phi (t)$, $\ddot \phi (t)$, $\dddot \phi (t)$, $H(t)$, and $\dot H(t)$
and does not depend on initial conditions of perturbation quantities. 
\section{Examples \label{sec4}}
As we have seen in the previous section, the expression of the effective growth rate function highly depends on the types of 
models. Therefore, we will consider three examples in this section; the quintessence model, kinetic gravity braiding model, and the case 
$G_4(\phi)=\mathrm{exp}[ \lambda \phi /M_{pl}]/(2 \kappa ^2)$, as 
the peculiar cases of $k$-essence model, $G_{3X} \neq 0$ model, and $G_{4 \phi} \neq 0$ model, respectively. 

\subsection{Quintessence model \label{secQ}}
In the case of quintessence model, i.e. $K(\phi ,X) = X-V(\phi)$, $G_{3}=0$, and $G_{4}=1/(2 \kappa ^2 )$, the 
effective growth rate function is described by 
\begin{equation}
f_\mathrm{eff} = \frac{\dot \phi }{H \phi}. 
\end{equation}
If we assume the slow-roll accelerated expansion, the conditions $| \ddot \phi | \ll 3H | \dot \phi | \simeq | V_\phi |$ and 
$3H^2 \simeq \kappa ^2 V$ are fulfilled. Then $f_\mathrm{eff}$ can be rewritten as 
\begin{equation}
f_\mathrm{eff} \simeq - \frac{V_{\phi}}{\kappa ^2 \phi V}. 
\label{slowfeff}
\end{equation}
Equation (\ref{slowfeff}) shows that the positive power law potential induces $f_\mathrm{eff}<0$ and the negative power law potential yields $f_\mathrm{eff} >0$. 
While, the quasi-static mode (\ref{eqqs}) usually has a growing solution, in other words, it has a positive growth factor. 
Therefore, we can ignore oscillating mode in the case of positive power law potential, on the other hand, in the case of negative power law potential, 
we should be careful with the behavior of the oscillating solutions. 
\begin{figure}
\begin{minipage}[t]{0.5\columnwidth}
\begin{center}
\includegraphics[clip, width=0.97\columnwidth]{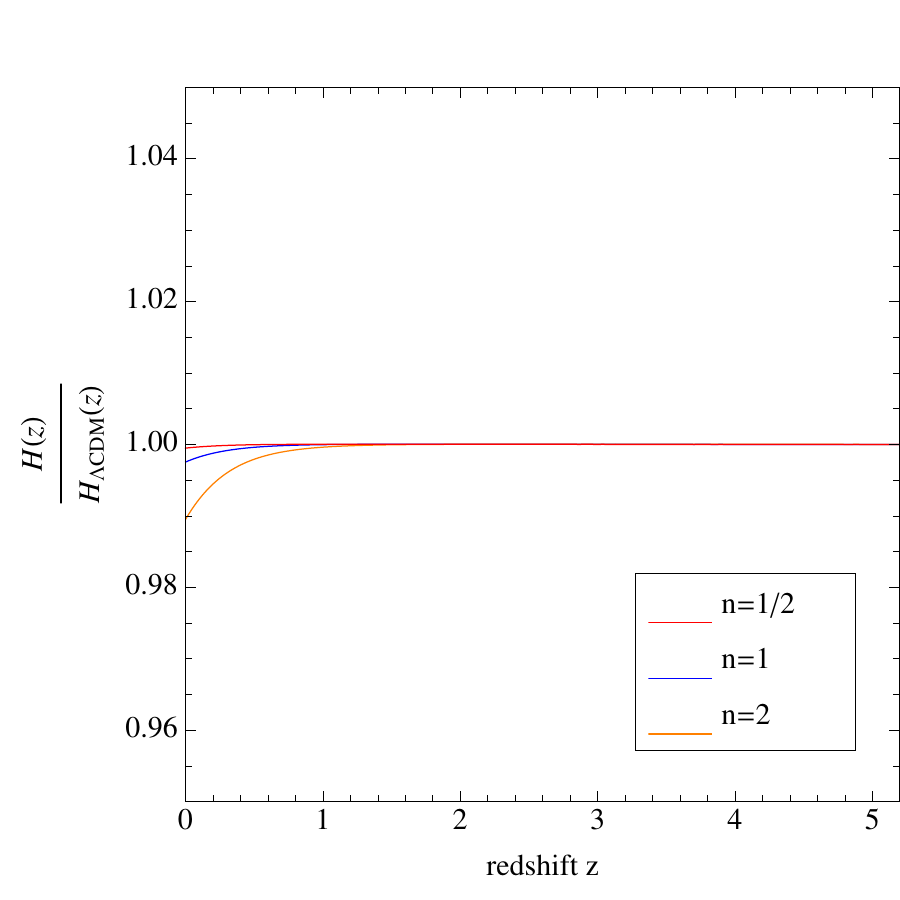}
\end{center}
\end{minipage}%
\begin{minipage}[t]{0.5\columnwidth}
\begin{center}
\includegraphics[clip, width=0.97\columnwidth]{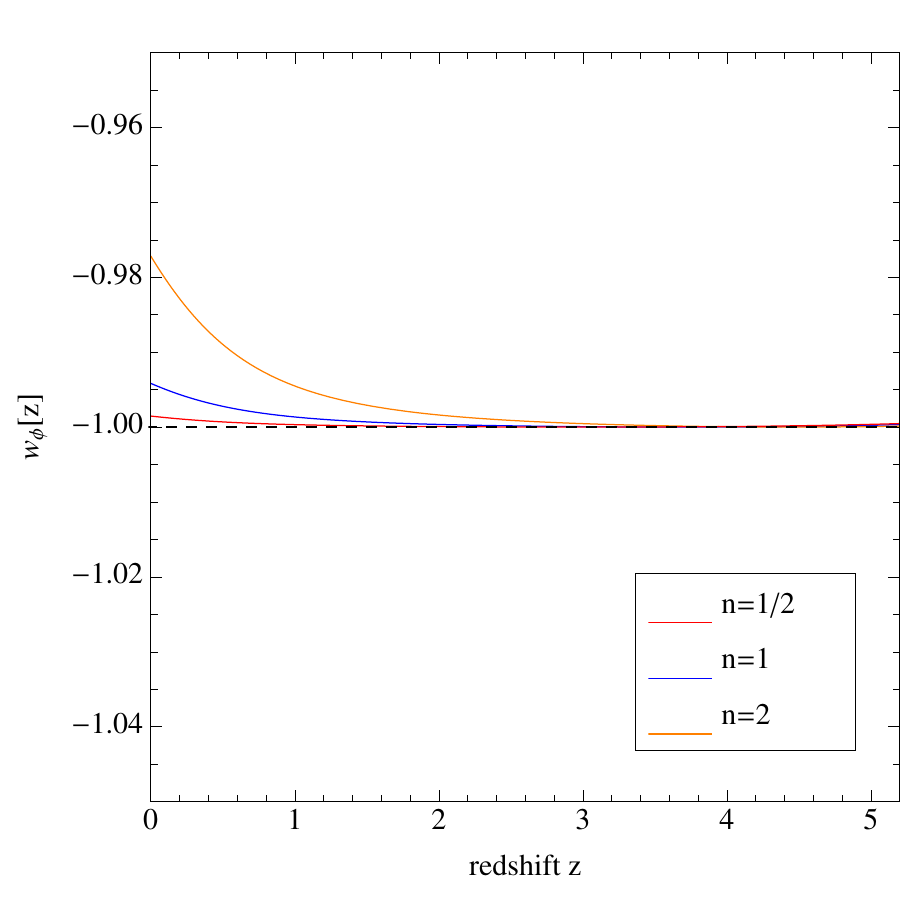}
\end{center}
\end{minipage}
\caption{Redshift dependence of the Hubble rate function compared to the $\Lambda$CDM model (left) and that of the effective 
equation of state parameter of the scalar field $w_\phi = p_\phi / \rho _\phi$ (right)
in the case of $K(\phi ,  X) = X- M^4 ( \phi /M_{pl}) ^{-n}$ , where $M^4 = 0.7 H_0^2 M_{pl}^2$ and 
$H_0$ means that the Hubble constant in the $\Lambda$CDM model, which is $H_0 \simeq 68$ (km/s)/Mpc.  
The initial conditions for $\phi (z)$ and $\dot \phi (z)$ are assigned as $\phi (10)=M_{pl}$ and $\dot \phi (10)=-0.04 M_{pl}H_0$,  
and also $\Omega _{\mathrm{matter},0}=0.31$ is assumed. 
}
\label{quint12}
\end{figure}
In Figs.~\ref{quint12} and \ref{quint34}, the behaviors of the background evolution and the growth rate functions in the case of 
negative power law potential are expressed.  
The background evolution is almost same as that of the $\Lambda$CDM model if we consider $V(\phi)=M^4 ( \phi /M_{pl}) ^{-n}$ with small $n$. 
The reason why only the cases $n \leq 2$ are plotted is that the astronomical observations and the experiments in solar system severely constrain the value of $n$ \cite{Farooq:2012ev,Matsumoto:2016lge}. 
While, the behavior of the growth rate function $f$ almost equivalents to that in the $\Lambda$CDM model 
for all $n$ as seen in Fig.~\ref{quint34} (left).  
 Figure \ref{quint34} (right) shows the effective growth rate function $f_\mathrm{eff}$ is always much smaller than the growth rate function $f$ in the case $n \leq 2$, 
 so we can justify to ignore the oscillating solutions. 

\begin{figure}
\begin{minipage}[t]{0.5\columnwidth}
\begin{center}
\includegraphics[clip, width=0.97\columnwidth]{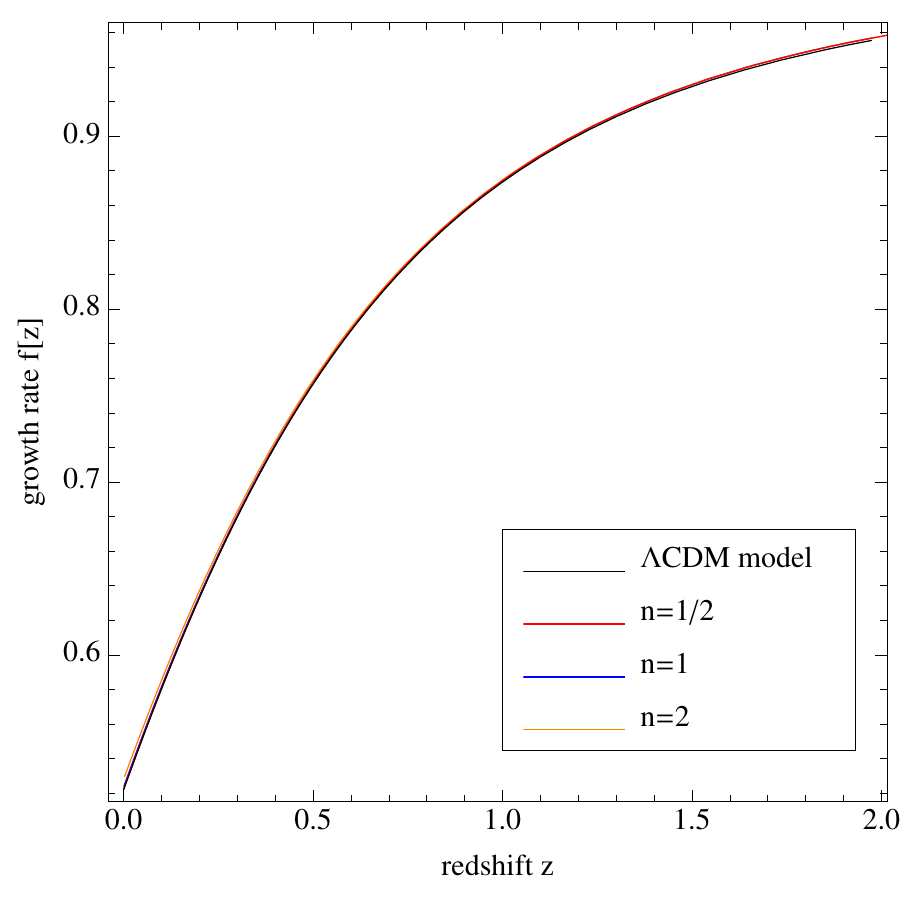}
\end{center}
\end{minipage}%
\begin{minipage}[t]{0.5\columnwidth}
\begin{center}
\includegraphics[clip, width=0.97\columnwidth]{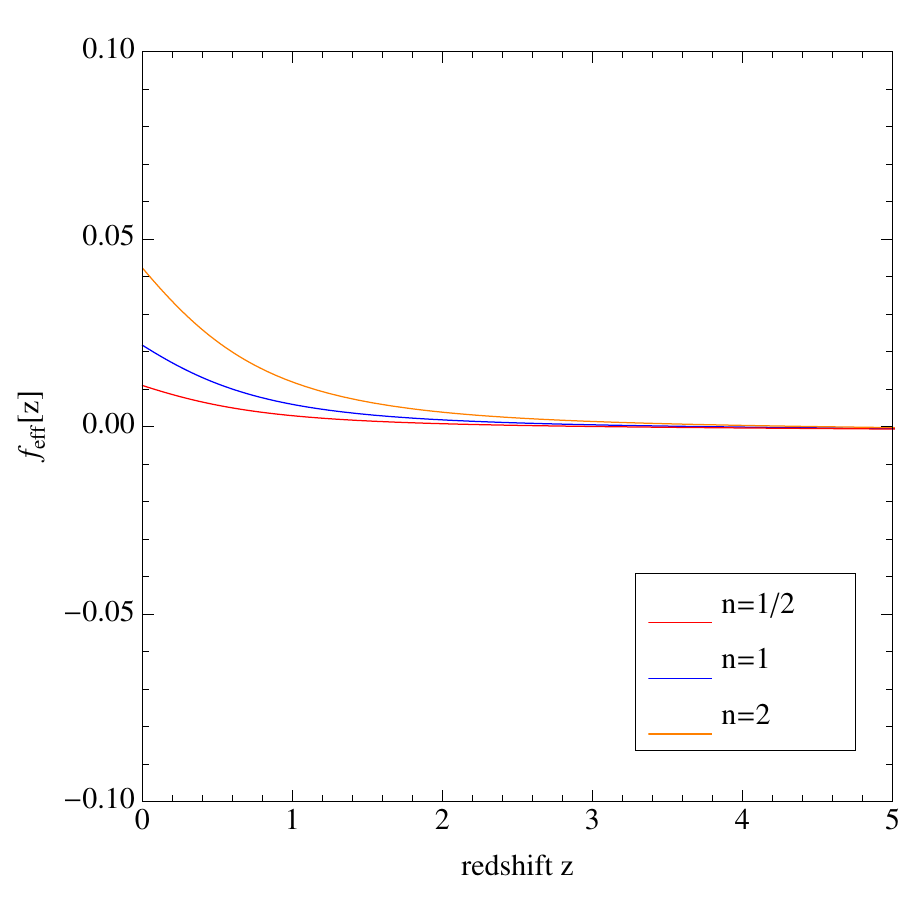}
\end{center}
\end{minipage}
\caption{Comparison between the growth rate function (left) and the 
effective growth rate function (right)
in the case of $K(\phi ,  X) = X- M^4 ( \phi /M_{pl}) ^{-n}$. 
Initial condition for the usual growth rate function $f$ is assigned as $f(z=10)=1$. }
\label{quint34}
\end{figure}
In the paper \cite{Bamba:2011ih}, the case of an exponential potential is investigated. 
In that case, the effective growth rate function becomes positive and can influence the 
total evolution of the matter density perturbation.  In particular, $f_\mathrm{eff} = 3/2 >1 =f $ 
is obtained for the limit $t \rightarrow 0$, therefore, we should take into account the behavior of the oscillating solutions. 
\subsection{Kinetic gravity braiding model}
\begin{figure}
\begin{minipage}[t]{0.5\columnwidth}
\begin{center}
\includegraphics[clip, width=0.97\columnwidth]{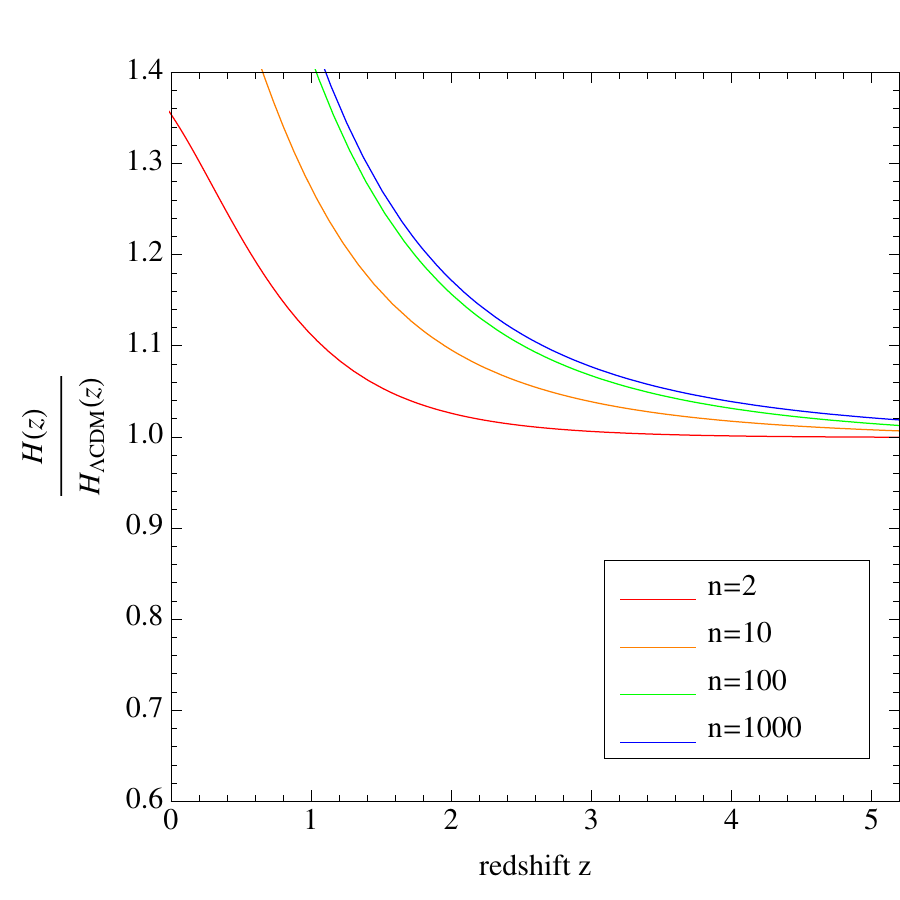}
\end{center}
\end{minipage}%
\begin{minipage}[t]{0.5\columnwidth}
\begin{center}
\includegraphics[clip, width=0.97\columnwidth]{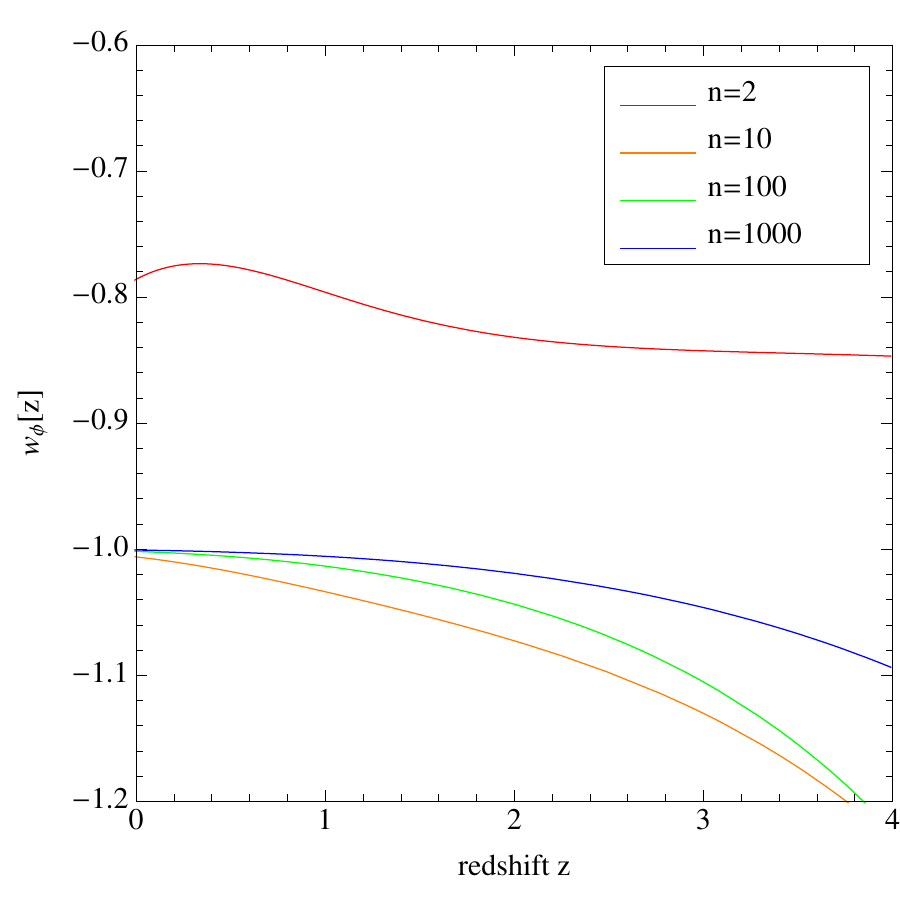}
\end{center}
\end{minipage}
\caption{Redshift dependence of the Hubble rate function (left) and that of 
equation of state parameter (right)
in the case of $K(\phi , X) = -X,$ $G_3(\phi , X) = M_{pl} ( r_c ^2 X/M_{pl}^2 ) ^n$ with $r_c = H_0 ^{-1}$. The 
initial condition for each curve is given by $\dot \phi _{red}(z=10) = 0.2M_{pl}H_0,$ $\dot \phi _{orange}(10)= 0.62M_{pl}H_0,$ $\dot \phi _{green}(10) = 0.94M_{pl}H_0,$ 
or $\dot \phi _{blue}(10) =0.995M_{pl}H_0$. 
}
\label{g3-1}
\end{figure}

\begin{figure}
\begin{center}
\includegraphics[clip, width=0.5\columnwidth]{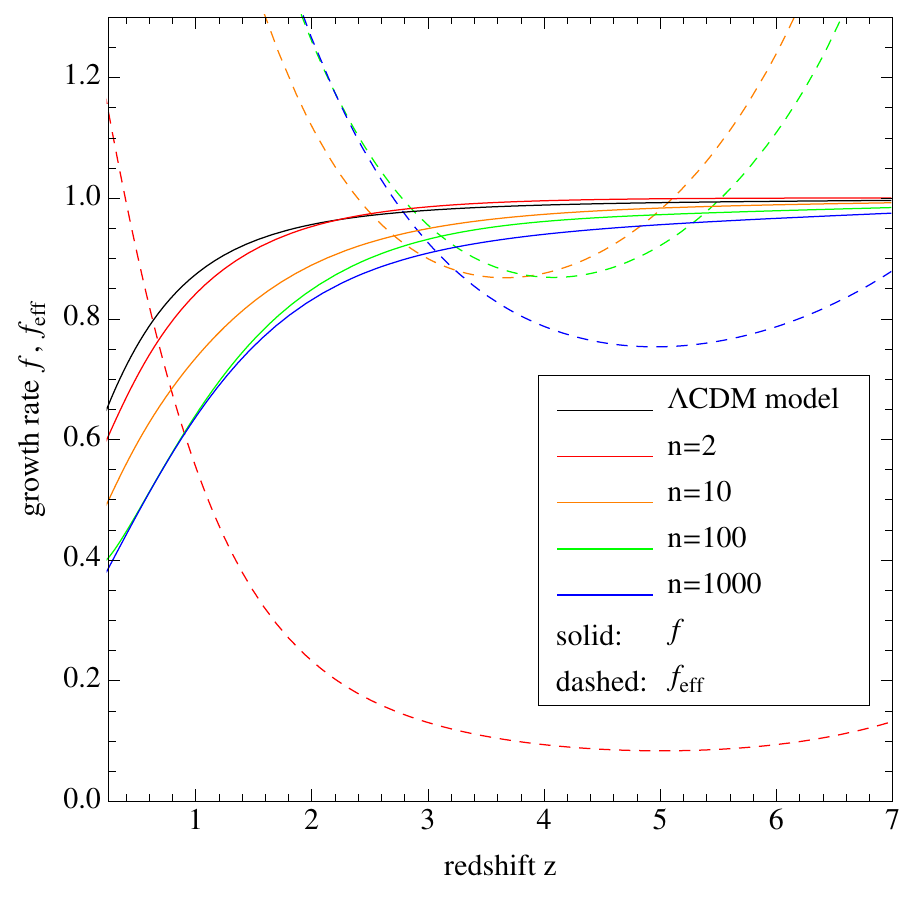}
\end{center}
\caption{Comparison between the growth rate function $f$ and the 
effective growth rate function $f_\mathrm{eff}$
in the case of $K(\phi , X) = -X,$ $G_3(\phi , X) = M_{pl} ( r_c ^2 X/M_{pl}^2 ) ^n$. The solid curves and the dashed curves express $f$ and $f_\mathrm{eff}$, respectively. 
Initial condition for the usual growth rate function $f$ is assigned as $f(z=10)=1$. }
\label{g3-2}
\end{figure}

Here, we consider kinetic gravity braiding model \cite{Deffayet:2010qz,Kimura:2010di}
\begin{equation}
K(\phi , X) = -X, \; G_3(\phi , X) = M_{pl} \left ( \frac{r_c ^2}{M_{pl}^2}X \right ) ^n, \; G_4(\phi)=\frac{1}{2 \kappa ^2}, 
\label{kgb}
\end{equation}
where $r_c$ and $n$ are positive constants. 
Then amplitude $C_{\delta}$ in Eq.~(\ref{cdeltagen}) is rewritten as 
\begin{align}
C_{\delta} \propto \frac{1}{a} G_{3X} \dot \phi ^2 \Big [ 8-24( n +1) G_{3X} H \dot \phi 
+3 G_{3X} ^2 \dot \phi ^2 \{ 6n(4n+5)H^2 - \kappa ^2 (3n+4 ) \dot \phi ^2  \}  \nonumber \\
+18 \kappa ^2 (n+2 ) G_{3X}^3H \dot \phi ^5 \Big ] 
\Big /  \Big [ 4-16 G_{3X} H \dot \phi  +2 G_{3X} ^2 \dot \phi ^2 \{ 30 H^2 -\kappa ^2 (3n+2 ) \dot \phi ^2 \} \nonumber  \\ 
-12 \kappa ^2 (n-2) G_{3X}^3 H \dot \phi ^5 -3 \kappa ^4 G_{3X}^4 \dot \phi ^8  \Big ] ^{5/4}, 
\end{align}
where we have eliminated the second derivative of $\phi$ and $\dot H$ by using Eqs.~(\ref{FL2}) and (\ref{FE}).
The effective growth rate function $f_\mathrm{eff}$ is obtained by differentiating $C_{\delta}$ with respect to $N$ and 
dividing it by $C_{\delta}$, however, the expression of $f_\mathrm{eff}$ is so messy that we do not explicitly show it here. 
In Fig.~\ref{g3-1}, the behaviors of the background evolution in the case of Eq.~(\ref{kgb}) are depicted. 
As shown in \cite{Kimura:2010di}, there are two attractors in this model; one has self-accelerating solution and  another do not have self-accelerating solution.
A large value of $\dot \phi$ is usually needed to realize the self-accelerating solution. 
Therefore, the results shown in Fig.~\ref{g3-1} does not so depend on the value of $\dot \phi (z=10)$ if it is enough large. 
In all the cases $n=2,10,100,1000$, the Hubble rate function is larger than that in the $\Lambda$CDM model in small redshift region, therefore, 
smaller growth rate $f$ than that in the $\Lambda$CDM model is realized as shown in Fig.~\ref{g3-2}. 
The effective growth rate function $f_\mathrm{eff}$ cannot be ignored in this case because 
$f_\mathrm{eff}$ becomes bigger than the growth rate $f$ in the region $z<2.5$ if $n =10,$ $100,$ or $1000$. 
In the case of $n=2$, the effective growth rate function could be ignored if the initial value 
of $f_\mathrm{eff}$ is much less than that of $f$. 
\subsection{The case $G_4(\phi)=\mathrm{exp}[ \lambda \phi /M_{pl}]/(2 \kappa ^2)$ and $G_3(\phi , X)=0$}
\begin{figure}
\begin{minipage}[t]{0.5\columnwidth}
\begin{center}
\includegraphics[clip, width=0.97\columnwidth]{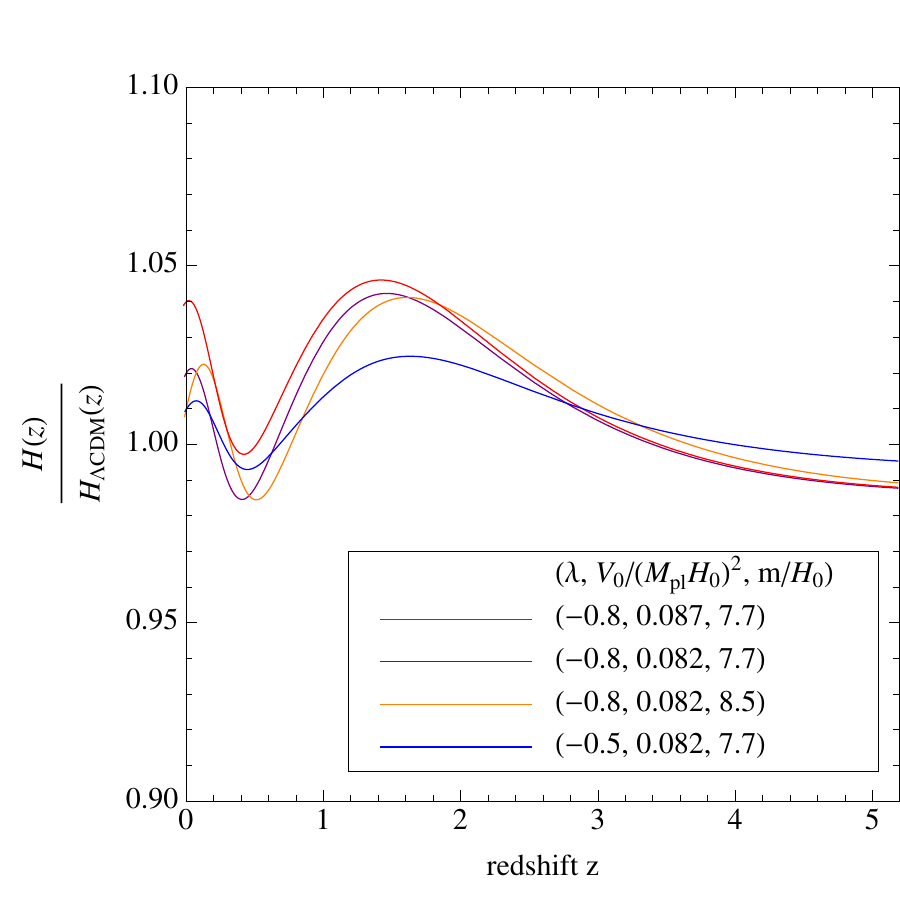}
\end{center}
\end{minipage}%
\begin{minipage}[t]{0.5\columnwidth}
\begin{center}
\includegraphics[clip, width=0.97\columnwidth]{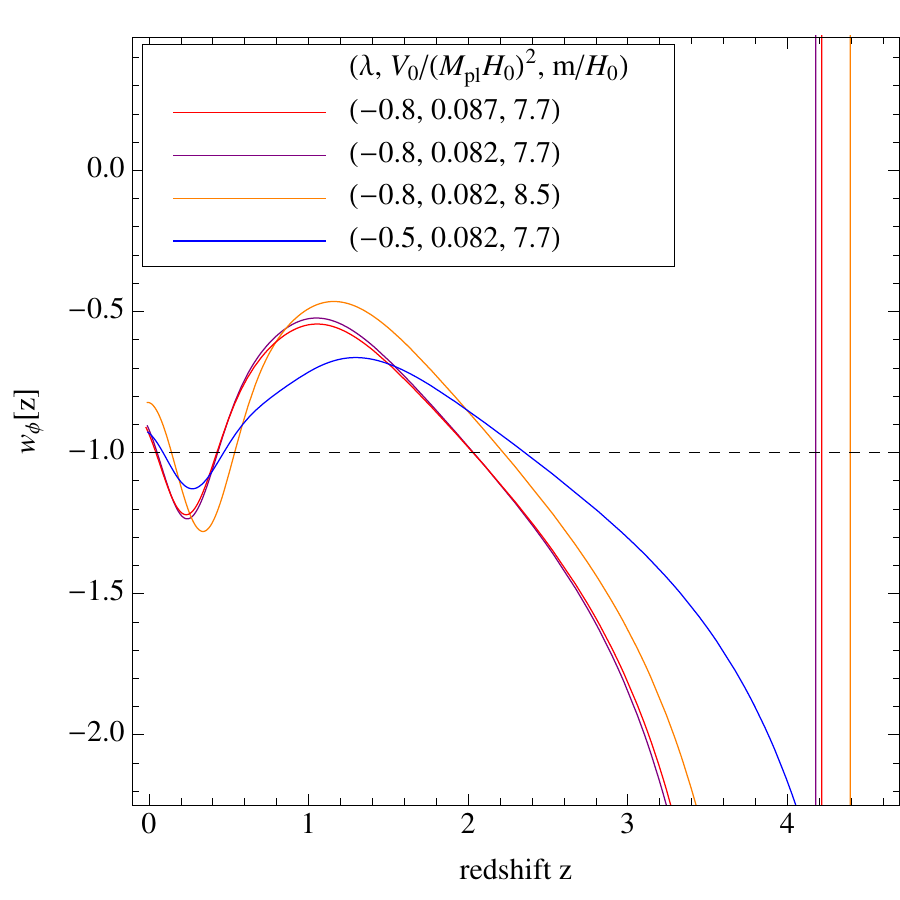}
\end{center}
\end{minipage}
\caption{Redshift dependence of Hubble rate function compared to that in the $\Lambda$CDM model (left) and that of effective equation of state parameter (right) 
in the case of $K(\phi ,  X) = X- (V_0+m^2 \phi ^2) $, $G_3(\phi , X)=0$, and $G_4(\phi )= \mathrm{exp}[ \lambda  \phi  / M_{pl}]/(2 \kappa ^2)$. 
}
\label{g4-1}
\end{figure}

\begin{figure}
\begin{minipage}[t]{0.5\columnwidth}
\begin{center}
\includegraphics[clip, width=0.97\columnwidth]{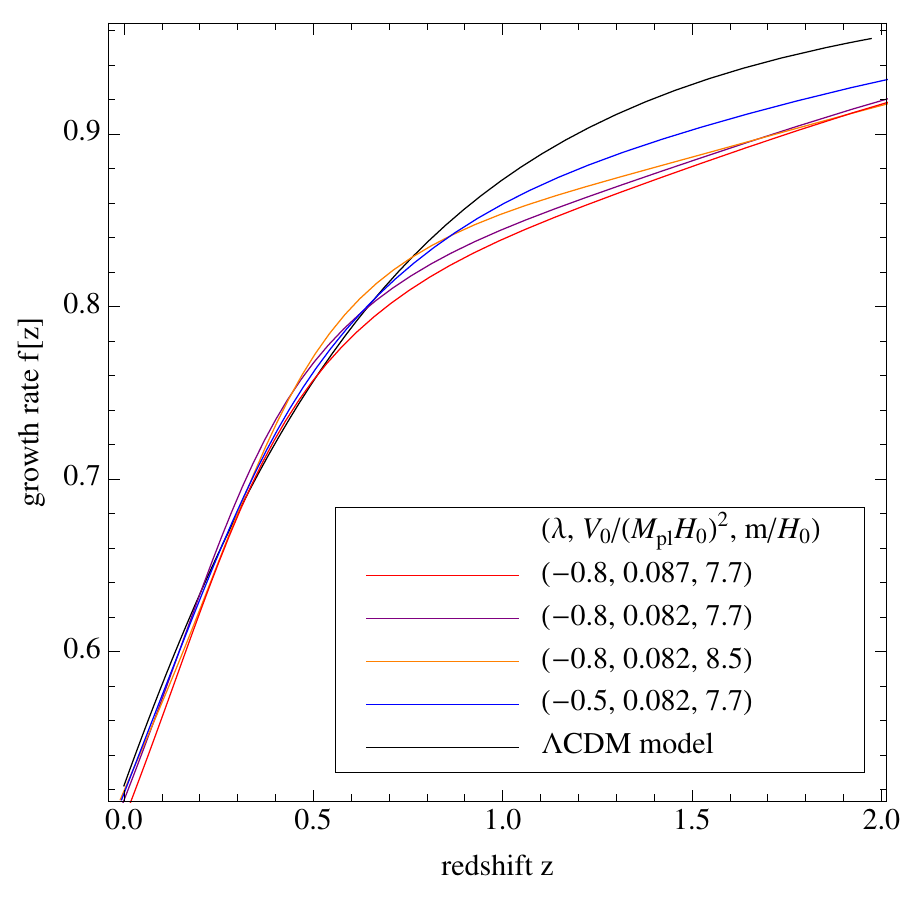}
\end{center}
\end{minipage}%
\begin{minipage}[t]{0.5\columnwidth}
\begin{center}
\includegraphics[clip, width=0.97\columnwidth]{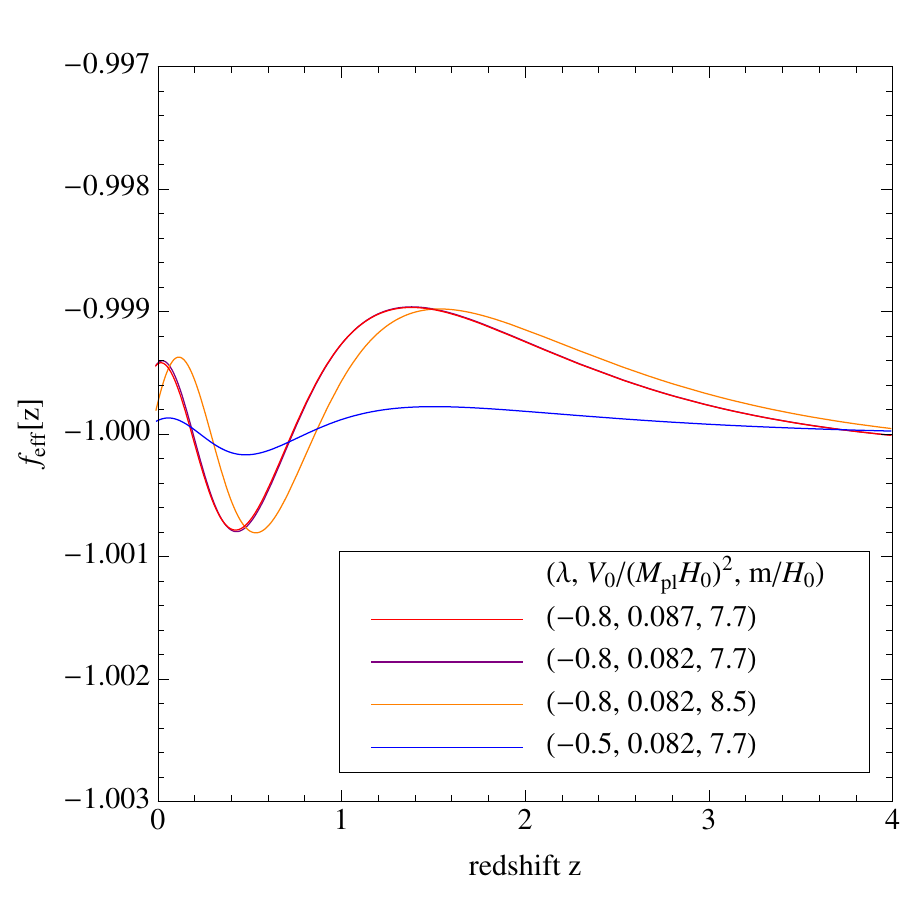}
\end{center}
\end{minipage}
\caption{Comparison between the growth rate function (left) and the 
effective growth rate function (right)
in the case of $K(\phi ,  X) = X- (V_0+m^2 \phi ^2) $, $G_3(\phi , X)=0$, and $G_4(\phi )= \mathrm{exp}[ \lambda  \phi  / M_{pl}]/(2 \kappa ^2)$. 
Initial condition for the usual growth rate function $f$ is assigned as $f(z=10)=1$. 
}
\label{g4-2}
\end{figure}

Let us consider the model 
\begin{equation}
K(\phi ,  X) = X- (V_0+m^2 \phi ^2) \quad and \quad G_4(\phi )= \frac{1}{2 \kappa ^2} \mathrm{e}^{\lambda  \frac{\phi }{ M_{pl}}}, \label{g4n2}
\end{equation}
which can realize phantom crossings without instability \cite{Matsumoto:2017qil}. Here, $m$ and $V_0$ are positive constants and $\lambda$ is a real constant. 
The background behavior of the model is shown in Fig.~\ref{g4-1}. In the right panel, we can see phantom crossings, which are 
the crossings of $w_\phi = -1$ line, are realized in the model. 
The effective growth function is expressed as 
\begin{align}
f_\mathrm{eff} = -1 - \frac{\lambda}{2} \frac{1}{1+ \frac{16 \pi}{3 \lambda ^2}\mathrm{e} ^{- \lambda \phi / M_{pl}}} \frac{\dot \phi}{M_{pl} H }. 
\end{align}
In the slow-roll regime, $| \dot \phi | / (M_{pl}H)$ is much smaller than $1$, so $f_ \mathrm{eff} \simeq -1$ is realized. 
In Fig.~\ref{g4-2}, the evolution of the growth rate $f$ and that of the effective growth rate $f_\mathrm{eff}$ are expressed. 
The behaviors of the growth rate function $f$ do not so depend on the values of the parameters except for the region $z>1$, because 
constant term $V_0$ dominate over the other terms. While, $f_ \mathrm{eff}$ is almost $-1$ all over the regime as we expected. 
That's why, the oscillation mode can be ignored in this case.

\section{Conclusion \label{sec5}}
In this paper, we have considered the behavior of the matter density perturbation in Horndeski's theory. 
The behavior of the quasi-static mode of the matter density perturbation is well known, however, 
that of oscillating mode has not been enough studied. 
Therefore, we have constructed the method to derive the oscillating solutions and have obtained the 
general expression of the solutions. 
In fact, oscillating solutions are the very features of dynamical dark energy models, 
because the existence of the extra degrees of freedom, which are scalar field perturbation and its time derivative, 
cause oscillating mode of the matter density perturbation. 
The peculiarity of the oscillating solutions are the following; 
the expression completely changes depending on the functions $G_3 (\phi , X)$ and $G_4 (\phi)$, 
amplitude of the solution depends not only on sound speed but also on higher derivatives of the functions, 
the oscillating mode can be a growing mode. 
The effective growth rate function of the oscillating mode is only expressed by background quantities and is independent from the 
initial conditions of the perturbation quantities.
In Sec.~\ref{sec4}, we have considered the cases, the quintessence model, kinetic gravity braiding model, and the case 
$G_4(\phi)=\mathrm{exp}[ \lambda \phi /M_{pl}]/(2 \kappa ^2)$, and have explicitly shown that there can be a growing and oscillating solution of the matter density perturbation. 
This result implies that oscillating mode of the matter density perturbation must be investigated for giving the true prediction of the models.  
The deviation between the quasi-static solution and the actual solution can be immediately evaluated by using Eq.~(\ref{cdeltak}) or (\ref{cdeltagen})
if we know background behavior of the model as we seen in Sec.~\ref{secQ}. 
While, the considerations of the concrete models of dark energy in Sec.~\ref{sec4} suggest that oscillating mode can be ignored if 
we assume the slow-roll accelerated expansion of the Universe. This subject will be studied near future. 
\section*{Acknowledgments}
The author's researches 
are supported by the Leung Center for Cosmology and Particle Astrophysics (LeCosPA) 
of the National Taiwan University (NTU). 
This work is partly supported by 
the Russian Government Program of Competitive Growth of Kazan Federal University.
\appendix
\section{Coefficients \label{app1}}
\begin{align}
A_1 =& -3 \dot \phi ^3 G_{3X} + 12HG_4 +6 \dot \phi G_{4 \phi}, \\
A_2 =& - \dot \phi (K_X + \dot \phi ^2 K_{XX}) + 2 \dot \phi G_{3 \phi} -3H \dot \phi ^2 (3 G_{3X} + \dot \phi ^2 G_{3XX}) + \dot \phi ^3 G_{3 \phi X} +6HG_{4 \phi}, \\
A_3 =& \; 4 G_4, \\
A_4 =& \; \dot \phi ^2 (K_X + \dot \phi ^2 K_{XX} ) -2 \dot \phi ^2 G_{3 \phi} - \dot \phi ^4 G_{3 \phi X} +3H \dot \phi ^3 (4 G_{3X} + \dot \phi ^2 G_{3XX}) \nonumber \\
&-12H(H G_4 + \dot \phi G_{4 \phi}), \\
A_5 =& \; \dot \phi ^3 G_{3X} -4H G_4 -2 \dot \phi G_{4 \phi}, \\
A_6 =& - \dot \phi ^2 G_{3X} +2 G_{4 \phi}, \\
\mu =& -K_\phi + \dot \phi ^2 K_{\phi X} - \dot \phi ^2 G_{3 \phi \phi} + 3H \dot \phi ^3 G_{3 \phi X} -6H^2 G_{4 \phi} -6H \dot \phi G_{4 \phi \phi}, \\
B_1 =& \; 12G_4, \\
B_2 =& -3 \dot \phi ^2 G_{3X} +6G_{4 \phi}, \\
B_3 =& \; 12(\dot \phi G_{4 \phi} + 3HG_4), \\
B_4 =& \; 3[ \dot \phi  K_X -2 \dot \phi G_{3 \phi} -2 \dot \phi \ddot \phi G_{3X} - \dot \phi ^3 (G_{3 \phi X} + \ddot \phi G_{3XX}) +4HG_{4 \phi} + 4 \dot \phi G_{4 \phi \phi}  ], \\
B_5 =& \; 3 (\dot \phi ^3 G_{3X} -4H G_4 -2 \dot \phi G_{4 \phi}), \\
B_6 =& \; 4G_4, \quad B_7 = 4 G_{4 \phi}, \quad B_8= 4G_4, \\
B_9 =& \; 3(2K- \dot \phi ^2 K_X +2 \dot \phi ^2 \ddot \phi G_{3X} + \dot \phi ^4 G_{3 \phi X} + \dot \phi ^4 \ddot \phi G_{3 XX} ), \\
\nu =& \; K_\phi - \dot \phi ^2 (G_{3 \phi \phi} + \ddot \phi G_{3 \phi X}) + 2 (3H^2 +2 \dot H)G_{4 \phi} +2 (\ddot \phi + 2H \dot \phi) G_{4 \phi \phi} + 2 \dot \phi ^2 G_{4 \phi \phi \phi}, \\
C_1 =& \; 4G_4, \\
C_2 =& - \dot \phi ^2  G_{3X} +2G_{4 \phi}, \\
C_3 =& \; \dot \phi ^3 G_{3X} -4HG_4 -2 \dot \phi G_{4 \phi}, \\
C_4 =& \; \dot \phi (K_X -2G_{3 \phi} +2 G_{4 \phi \phi} ) +H(3 \dot \phi ^2 G_{3X} -2G_{4 \phi }), \\
D_1 =& -3( \dot \phi ^2 G_{3X} -2 G_{4 \phi}), \\
D_2 =& -K_X - \dot \phi ^2 K_{XX} +2G_{3 \phi} -6H \dot \phi G_{3X} + \dot \phi ^2 G_{3 \phi X} -3H \dot \phi ^3 G_{3XX}, \\
D_3 =& -3(\dot \phi K_X -2 \dot \phi G_{3 \phi} + 6H \dot \phi ^2 G_{3X} +2 \dot \phi \ddot \phi G_{3X} + \dot \phi ^3 G_{3 \phi X} + \dot \phi ^3 \ddot \phi G_{3XX} -8H G_{4 \phi}), \\
D_4 =& \; \frac{d}{dt} D_2 +3HD_2, \\
D_5 =& \; \dot \phi (K_X + \dot \phi ^2 K_{XX} -2G_{3 \phi} - \dot \phi ^2 G_{3 \phi X}) + 3H(3 \dot \phi ^2 G_{3X} +  \dot \phi ^4 G_{3XX}-2 G_{4 \phi}), \\
D_7 =& \; 4 G_{4 \phi},  \\
D_8 =& \; 9H \dot \phi ^{-1}K + 3K_\phi -3 (\ddot \phi + 3H \dot \phi )K_X -3 \dot \phi ^2 (K_{\phi X}+ \ddot \phi K_{XX}) \nonumber \\
&+ 3( 2 \ddot \phi +3H \dot \phi ) G_{3 \phi} -9 \dot \phi (3H \ddot \phi + 3H^2 \dot \phi + \dot H \dot \phi ) G_{3X} +3 \dot \phi ^2 G_{3 \phi \phi} \nonumber \\
&+ 3 \dot \phi ^2 (\ddot \phi -3H \dot \phi ) G_{3 \phi X} -9H \dot \phi ^3 \ddot \phi G_{3XX} + 18H \dot \phi ^{-1} (3H^2 + 2 \dot H) G_4 \nonumber \\
&+ 18 \dot \phi ^{-1}(H \ddot \phi + 4H^2 \dot \phi + \dot H \dot \phi ) G_{4 \phi} +18 H \dot \phi G_{4 \phi \phi}, \\
D_9 =& -K_X+2G_{3 \phi} -4H \dot \phi G_{3X} - \ddot \phi (2G_{3X} + \dot \phi ^2 G_{3XX}) - \dot \phi ^2 G_{3 \phi X}, \\
D_{10} =& - \dot \phi ^2 G_{3X} + 2G_{4 \phi}, \\
D_{11} =& \; K_\phi + ( \ddot \phi +3H \dot \phi ) K_X + \dot \phi ^2 (4 \ddot \phi + 3H \dot \phi ) K_{XX} +\dot \phi ^4 (K_{\phi XX} + \ddot \phi K_{XXX}) \nonumber \\
& -2 (\ddot \phi + 3H \dot \phi ) G_{3 \phi} +9 \dot \phi (2H \ddot \phi +3H^2 \dot \phi + \dot H \dot \phi) G_{3X} - \dot \phi ^2 G_{3 \phi \phi} \nonumber \\
&- \dot \phi ^2 (5 \ddot \phi -3H \dot \phi ) G_{3 \phi X} +3 \dot \phi ^3 (7H \ddot \phi +3H^2 \dot \phi + \dot H \dot \phi ) G_{3XX} - \dot \phi ^4 G_{3 \phi \phi X} \nonumber \\
&- \dot \phi ^4 (\ddot \phi -3H \dot \phi) G_{3 \phi XX} + 3H \dot \phi ^5 \ddot \phi G_{3XXX} -6 (2H^2 + \dot H) G_{4 \phi}, \\
M^2 =& -K_{\phi \phi} +(\ddot \phi + 3H \dot \phi)K_{\phi X} + \dot \phi ^2 K_{\phi \phi X} + \dot \phi ^2 \ddot \phi K_{\phi XX} - \ddot \phi [ 2G_{3 \phi \phi} + \dot \phi ^2 G_{3 \phi \phi X} \nonumber \\
&-3H \dot \phi (2G_{3 \phi X} + \dot \phi ^2 G_{3 \phi XX} )] -6H \dot \phi G_{3 \phi \phi} + 3 \dot \phi ^2 (3H^2 + \dot H) G_{3 \phi X} - \dot \phi ^2 G_{3 \phi \phi \phi} \nonumber \\
&+3H \dot \phi ^3 G_{3 \phi \phi X} -6(2H^2 + \dot H)G_{4 \phi \phi}, 
\end{align}


\begin{thebibliography}{99}
\bibitem{Riess:1998cb}
  A.~G.~Riess {\it et al.} [Supernova Search Team Collaboration],
  Astron.\ J.\  {\bf 116}, 1009 (1998)
  [astro-ph/9805201].

\bibitem{Perlmutter:1998np}
  S.~Perlmutter {\it et al.} [Supernova Cosmology Project Collaboration],
  Astrophys.\ J.\  {\bf 517}, 565 (1999)
  [astro-ph/9812133].

\bibitem{Komatsu:2010fb}
  E.~Komatsu {\it et al.} [WMAP Collaboration],
  Astrophys.\ J.\ Suppl.\  {\bf 192}, 18 (2011)
  [arXiv:1001.4538 [astro-ph.CO]].

\bibitem{Ade:2013zuv}
  P.~A.~R.~Ade {\it et al.} [Planck Collaboration],
  Astron.\ Astrophys.\  {\bf 571}, A16 (2014)
  [arXiv:1303.5076 [astro-ph.CO]].

\bibitem{Ade:2015xua} 
  P.~A.~R.~Ade {\it et al.} [Planck Collaboration],
  Astron.\ Astrophys.\  {\bf 594}, A13 (2016)
  [arXiv:1502.01589 [astro-ph.CO]].

\bibitem{Percival:2009xn}
  W.~J.~Percival {\it et al.} [SDSS Collaboration],
  Mon.\ Not.\ Roy.\ Astron.\ Soc.\  {\bf 401}, 2148 (2010)
  [arXiv:0907.1660 [astro-ph.CO]].

\bibitem{Blake:2011en}
  C.~Blake {\it et al.},
  Mon.\ Not.\ Roy.\ Astron.\ Soc.\  {\bf 418}, 1707 (2011)
  [arXiv:1108.2635 [astro-ph.CO]].

\bibitem{Beutler:2011hx}
  F.~Beutler {\it et al.},
  Mon.\ Not.\ Roy.\ Astron.\ Soc.\  {\bf 416}, 3017 (2011)
  [arXiv:1106.3366 [astro-ph.CO]].

\bibitem{Cuesta:2015mqa}
  A.~J.~Cuesta {\it et al.},
  Mon.\ Not.\ Roy.\ Astron.\ Soc.\  {\bf 457}, no. 2, 1770 (2016)
  [arXiv:1509.06371 [astro-ph.CO]].

\bibitem{Delubac:2014aqe}
  T.~Delubac {\it et al.} [BOSS Collaboration],
  Astron.\ Astrophys.\  {\bf 574}, A59 (2015)
  [arXiv:1404.1801 [astro-ph.CO]]. 
  
\bibitem{Peebles:1987ek} 
  P.~J.~E.~Peebles and B.~Ratra,
  Astrophys.\ J.\  {\bf 325}, L17 (1988).

\bibitem{Ratra:1987rm} 
  B.~Ratra and P.~J.~E.~Peebles,
  Phys.\ Rev.\ D {\bf 37}, 3406 (1988).

\bibitem{Chiba:1997ej} 
  T.~Chiba, N.~Sugiyama and T.~Nakamura,
  Mon.\ Not.\ Roy.\ Astron.\ Soc.\  {\bf 289}, L5 (1997)
  [astro-ph/9704199].

\bibitem{Zlatev:1998tr} 
  I.~Zlatev, L.~M.~Wang and P.~J.~Steinhardt,
  Phys.\ Rev.\ Lett.\  {\bf 82}, 896 (1999)
  [astro-ph/9807002]. 
  

\bibitem{Jordan:1955}
P.~Jordan, ``Schwerkraft und Weltall,'' (Fredrich Vieweg und Sohn, Brunschweig, 1955). 

\bibitem{Brans:1961sx} 
  C.~Brans and R.~H.~Dicke,
  Phys.\ Rev.\  {\bf 124}, 925 (1961).
  
\bibitem{Fujii:2003pa} 
  Y.~Fujii and K.~Maeda,
  ``The scalar-tensor theory of gravitation,'' 
  (Cambridge Monographs on Mathematical Physics, Cambridge Univ. Press, 
  Cambridge, United Kingdom, 2003). 
  
\bibitem{Maeda:1988ab} 
  K.~i.~Maeda,
  Phys.\ Rev.\ D {\bf 39}, 3159 (1989).

\bibitem{Buchdahl:1983zz} 
  H.~A.~Buchdahl,
  Mon.\ Not.\ Roy.\ Astron.\ Soc.\  {\bf 150}, 1 (1970).
  
\bibitem{Nojiri:2006ri} 
  S.~Nojiri and S.~D.~Odintsov,
  eConf C {\bf 0602061}, 06 (2006)
  [Int.\ J.\ Geom.\ Meth.\ Mod.\ Phys.\  {\bf 4}, 115 (2007)]
  [hep-th/0601213].
  
\bibitem{Sotiriou:2008rp} 
  T.~P.~Sotiriou and V.~Faraoni,
  Rev.\ Mod.\ Phys.\  {\bf 82}, 451 (2010)
  [arXiv:0805.1726 [gr-qc]].
  
\bibitem{DeFelice:2010aj} 
  A.~De Felice and S.~Tsujikawa,
  Living Rev.\ Rel.\  {\bf 13}, 3 (2010)
  [arXiv:1002.4928 [gr-qc]].

\bibitem{Nojiri:2010wj} 
  S.~Nojiri and S.~D.~Odintsov,
  Phys.\ Rept.\  {\bf 505}, 59 (2011)
  [arXiv:1011.0544 [gr-qc]].
  
\bibitem{Horndeski}
G. W. Horndeski,
  Int. J. Theor. Phys. 10, 363-384 (1974). 

\bibitem{Huterer:1998qv} 
  D.~Huterer and M.~S.~Turner,
  Phys.\ Rev.\ D {\bf 60}, 081301 (1999)
  [astro-ph/9808133].
  
\bibitem{Starobinsky:1998fr} 
  A.~A.~Starobinsky,
  JETP Lett.\  {\bf 68}, 757 (1998)
  [Pisma Zh.\ Eksp.\ Teor.\ Fiz.\  {\bf 68}, 721 (1998)]
  [astro-ph/9810431].

\bibitem{Tsujikawa:2007gd} 
  S.~Tsujikawa,
  Phys.\ Rev.\ D {\bf 76}, 023514 (2007)
  [arXiv:0705.1032 [astro-ph]].
  
\bibitem{delaCruzDombriz:2008cp} 
  A.~de la Cruz-Dombriz, A.~Dobado and A.~L.~Maroto,
  Phys.\ Rev.\ D {\bf 77}, 123515 (2008)
  [arXiv:0802.2999 [astro-ph]].

\bibitem{DeFelice:2011hq} 
  A.~De Felice, T.~Kobayashi and S.~Tsujikawa,
  Phys.\ Lett.\ B {\bf 706}, 123 (2011)
  [arXiv:1108.4242 [gr-qc]].
  
\bibitem{Bamba:2011ih} 
  K.~Bamba, J.~Matsumoto and S.~Nojiri,
  Phys.\ Rev.\ D {\bf 85}, 084026 (2012)
  [arXiv:1109.1308 [hep-th]].
 
\bibitem{Matsumoto:2013sba} 
  J.~Matsumoto,
  Phys.\ Rev.\ D {\bf 87}, no. 10, 104002 (2013)
  [arXiv:1303.6828 [hep-th]].
  
\bibitem{Kobayashi:2009wr} 
  T.~Kobayashi, H.~Tashiro and D.~Suzuki,
  Phys.\ Rev.\ D {\bf 81}, 063513 (2010)
  [arXiv:0912.4641 [astro-ph.CO]].
  
\bibitem{Deffayet:2011gz} 
  C.~Deffayet, X.~Gao, D.~A.~Steer and G.~Zahariade,
  Phys.\ Rev.\ D {\bf 84}, 064039 (2011)
  [arXiv:1103.3260 [hep-th]].
  
\bibitem{Kobayashi:2011nu} 
  T.~Kobayashi, M.~Yamaguchi and J.~Yokoyama,
  Prog.\ Theor.\ Phys.\  {\bf 126}, 511 (2011)
  [arXiv:1105.5723 [hep-th]].
  
\bibitem{DeFelice:2011bh} 
  A.~De Felice and S.~Tsujikawa,
  JCAP {\bf 1202}, 007 (2012)
  [arXiv:1110.3878 [gr-qc]].
  
\bibitem{TheLIGOScientific:2017qsa} 
  B.~P.~Abbott {\it et al.} [LIGO Scientific and Virgo Collaborations],
  Phys.\ Rev.\ Lett.\  {\bf 119}, no. 16, 161101 (2017)
  [arXiv:1710.05832 [gr-qc]].
 
\bibitem{Monitor:2017mdv} 
  B.~P.~Abbott {\it et al.}, 
  Astrophys.\ J.\  {\bf 848}, no. 2, L13 (2017)
  [arXiv:1710.05834 [astro-ph.HE]].
 
\bibitem{GBM:2017lvd} 
  B.~P.~Abbott {\it et al.}, 
  Astrophys.\ J.\  {\bf 848}, no. 2, L12 (2017)
  [arXiv:1710.05833 [astro-ph.HE]]. 
  
\bibitem{Coulter:2017wya} 
  D.~A.~Coulter {\it et al.},
  Science
  [Science {\bf 358}, 1556 (2017)]
  [arXiv:1710.05452 [astro-ph.HE]].
  
\bibitem{Creminelli:2017sry} 
  P.~Creminelli and F.~Vernizzi,
  Phys.\ Rev.\ Lett.\  {\bf 119}, no. 25, 251302 (2017)
  [arXiv:1710.05877 [astro-ph.CO]].
  
\bibitem{Sakstein:2017xjx} 
  J.~Sakstein and B.~Jain,
  Phys.\ Rev.\ Lett.\  {\bf 119}, no. 25, 251303 (2017)
  [arXiv:1710.05893 [astro-ph.CO]].

\bibitem{Ezquiaga:2017ekz} 
  J.~M.~Ezquiaga and M.~Zumalacárregui,
  Phys.\ Rev.\ Lett.\  {\bf 119}, no. 25, 251304 (2017)
  [arXiv:1710.05901 [astro-ph.CO]].
  
\bibitem{Baker:2017hug} 
  T.~Baker, E.~Bellini, P.~G.~Ferreira, M.~Lagos, J.~Noller and I.~Sawicki,
  Phys.\ Rev.\ Lett.\  {\bf 119}, no. 25, 251301 (2017)
  [arXiv:1710.06394 [astro-ph.CO]].
  
\bibitem{Arai:2017hxj} 
  S.~Arai and A.~Nishizawa,
  Phys.\ Rev.\ D {\bf 97}, no. 10, 104038 (2018)
  [arXiv:1711.03776 [gr-qc]].
  
\bibitem{Langlois:2017dyl} 
  D.~Langlois, R.~Saito, D.~Yamauchi and K.~Noui,
  arXiv:1711.07403 [gr-qc]. 

\bibitem{Lombriser:2015sxa} 
  L.~Lombriser and A.~Taylor,
  JCAP {\bf 1603}, no. 03, 031 (2016)
  [arXiv:1509.08458 [astro-ph.CO]]. 
  
\bibitem{Lombriser:2016yzn} 
  L.~Lombriser and N.~A.~Lima,
  Phys.\ Lett.\ B {\bf 765}, 382 (2017)
  [arXiv:1602.07670 [astro-ph.CO]].

\bibitem{ArmendarizPicon:1999rj} 
  C.~Armendariz-Picon, T.~Damour and V.~F.~Mukhanov,
  Phys.\ Lett.\ B {\bf 458}, 209 (1999)
  [hep-th/9904075].
  
\bibitem{Garriga:1999vw} 
  J.~Garriga and V.~F.~Mukhanov,
  Phys.\ Lett.\ B {\bf 458}, 219 (1999)
  [hep-th/9904176].
 
\bibitem{ArmendarizPicon:2000ah} 
  C.~Armendariz-Picon, V.~F.~Mukhanov and P.~J.~Steinhardt,
  Phys.\ Rev.\ D {\bf 63}, 103510 (2001)
  [astro-ph/0006373].
  
\bibitem{Farooq:2012ev} 
  O.~Farooq, D.~Mania and B.~Ratra,
  Astrophys.\ J.\  {\bf 764}, 138 (2013)
  [arXiv:1211.4253 [astro-ph.CO]].

\bibitem{Matsumoto:2016lge} 
  J.~Matsumoto,
  Int.\ J.\ Mod.\ Phys.\ D {\bf 27}, no. 01, 1750173 (2017)
  [arXiv:1607.03003 [gr-qc]].  
 
\bibitem{Deffayet:2010qz} 
  C.~Deffayet, O.~Pujolas, I.~Sawicki and A.~Vikman,
  JCAP {\bf 1010}, 026 (2010)
  [arXiv:1008.0048 [hep-th]].
  
\bibitem{Kimura:2010di} 
  R.~Kimura and K.~Yamamoto,
  JCAP {\bf 1104}, 025 (2011)
  [arXiv:1011.2006 [astro-ph.CO]].
  
\bibitem{Matsumoto:2017qil} 
  J.~Matsumoto,
  Phys.\ Rev.\ D {\bf 97}, no. 12, 123538 (2018)
  [arXiv:1712.10015 [gr-qc]].
\end{thebibliography}
\end{document}